\documentclass[a4paper,12pt]{article}
\usepackage{setspace}
\usepackage[top=2.54cm,bottom=2.54cm, left=2.2cm, right=2.2cm]{geometry}
\usepackage{bm}
\usepackage[longnamesfirst]{natbib}
\usepackage{graphicx}
\usepackage{epstopdf}
\usepackage{amsmath}
\usepackage{hyperref}
\usepackage{doi}

\begin{document}

\begin{center}
\Large{The $k$-NN algorithm for compositional data: a revised approach with and without zero values present}
\\
\normalsize
\bigskip \bigskip
{\bf Michail Tsagris} \\
School of Mathematical Sciences, University of Nottingham, UK \\
\href{mailto:mtsagris@yahoo.gr}{mtsagris@yahoo.gr}, 
\end{center}

\begin{center}
{\bf Abstract}
\end{center}
In compositional data, an observation is a vector with non-negative components which sum to a constant, typically 1. Data of this type arise in many areas, such as geology, archaeology, biology, economics and political science among others. The goal of this paper is to extend the taxicab metric and a newly suggested metric for compositional data by employing a power transformation. Both  metrics are to be used in the $k$-nearest neighbours algorithm regardless of the presence of zeros. Examples with real data are exhibited. 

\section{Introduction}
Compositional data are non-negative multivariate data and each vector sums to the same constant, usually $1$ for convenience. Compositional data are met in many disciplines, including geology \citep{ait1982}, economics \citep*{fry2000}, archaeology \citep{baxter2005} and political sciences \citep{rodrigues2009}. Their sample space is called simplex $S^d$ and in mathematical terms is 
\begin{footnotesize}
\begin{eqnarray*}
S^d=\left\lbrace(x_1,...,x_D)^T \bigg\vert x_i \geq 0,\sum_{i=1}^Dx_i=1\right\rbrace, 
\end{eqnarray*}
\end{footnotesize}
where $D$ denotes the number of components and $d=D-1$.

Ever since \citet{ait1982} suggested the use of the log-ratio transformation for compositional data, most of the analyses of such data have been implemented using this transformation. \citet{ait2003} implemented linear discriminant analysis for compositional data using the log-ratio transformation. Over the years though, researchers have suggested alternative ways for supervised classification of compositional data, see for example \citet{gallo2010} and \citet{neocleous2011}. 

An important issue in compositional data is the presence of zeros, which cause problems for the logarithmic transformation. The issue of zero values in some components is not addressed in most papers, but see \citet{neocleous2011} for an example of discrimination in the presence of zeros. Alternatively, one could use alternative models (see for example \citealp{scealy2011a} and \citealp{stewart2011}) or replace the zero values by making parametric assumptions \citep{martin2012}. 

In this paper we suggest the use of a recently developed metric, for classification of compositional data, when the $k$-nearest neighbours ($k$-NN) algorithm is implemented. It is a metric for probability distributions \citep{endres2003, osterreicher2003} which can be adopted to compositional data as well, since each vector sums to $1$. The second metric we suggest is the Manhattan metric, a scaled version of which has already been used for compositional data analysis \citep{miller2002}. We will extend both of these metrics by applying a power transformation. We will see that both of these metrics handle zeros naturally and hence they can be used regardless of them being present in some components. This is a very attractive feature of these metrics in contrast to the Aitchisonian metric suggested by \citet{ait2003} which is not applicable when zeros are present in the data. Examples using real data are used to illustrate the performance of these metrics. 

Section 2 describes the two metrics, how they can be extended and also presents graphically their loci of points equidistant from the centre of the simplex. Section 3 shows the $k$-NN algorithm for compositional data and Section 4 contains examples using real data. Finally, section 5 concludes this paper.

\section{Metrics for compositional data}
We will present three metrics for compositional data ,two of which have already been examined. But first we will show the power transformation.
\citet{ait2003} defined the power transformation to be
\begin{footnotesize}
\begin{eqnarray} \label{alpha}
{\bf u}=\left( \frac{x_1^{\alpha}}{\sum_{j=1}^Dx_j^{\alpha}}, \ldots, \frac{x_D^{\alpha}}{\sum_{j=1}^Dx_j^{\alpha}} \right)^T.
\end{eqnarray}
\end{footnotesize}
The value of $\alpha$ will be determined by the estimated accuracy of the $k$-NN algorithm. 

\subsection{The ES-OV$_\alpha$ metric for compositional data}
We advocate that as a measure of the distance between two compositions we can use the square root of the Jensen-Shannon divergence
\begin{footnotesize}
\begin{eqnarray} \label{ESOV}
ES-OV({\bf x},{\bf w})=\left[ \sum_{i=1}^D\left( x_i\log{\frac{2x_i}{x_i+w_i}}+w_i\log{\frac{2w_i}{x_i+w_i}} \right) \right]^{1/2},
\end{eqnarray}
\end{footnotesize}
where  ${\bf x}, {\bf w} \in S^d$. 
 
\citet{endres2003} and \citet{osterreicher2003} proved, independently, that (\ref{ESOV}) satisfies the triangular identity and thus it is a metric. For this reason we will refer to it as the ES-OV metric. 

We will use the power transformation (\ref{alpha}) to define a more general metric termed ES-OV$_{\alpha}$ metric 
\begin{footnotesize}
\begin{eqnarray} \label{ESOVa}
ES-OV_{\alpha}({\bf x},{\bf w})=\left[ \sum_{i=1}^D\left( \frac{x^{\alpha}_i}{\sum_{j=1}^Dx^{\alpha}_j}
\log{\frac{2\frac{x^{\alpha}_i}{\sum_{j=1}^Dx^{\alpha}_j}}{\frac{x^{\alpha}_i}{\sum_{j=1}^Dx^{\alpha}_j}+
\frac{w^{\alpha}_i}{\sum_{j=1}^Dw^{\alpha}_j}}}+
\frac{w^{\alpha}_i}{\sum_{j=1}^Dw^{\alpha}_j}\log{\frac{2\frac{w^{\alpha}_i}{\sum_{j=1}^Dw^{\alpha}_j}}{\frac{x^{\alpha}_i}{\sum_{j=1}^Dx^{\alpha}_j}+\frac{w^{\alpha}_i}{\sum_{j=1}^Dw^{\alpha}_j}}} \right) \right]^{1/2}.
\end{eqnarray}
\end{footnotesize}

\subsection{The taxicab$_{\alpha}$ metric for compositional data}
The taxicab metric is also known as $L_1$ (or Manhattan) metric and is defined as
\begin{footnotesize}
\begin{eqnarray} \label{taxicab}
TC\left({\bf x},{\bf w}\right)=\sum_{i=1}^D\left| x_i-w_i \right|
\end{eqnarray}
\end{footnotesize}
We will again employ the power transformation (\ref{alpha}) to define a more general metric which we will term the TC$_{\alpha}$ metric
\begin{footnotesize}
\begin{eqnarray} \label{taxicaba}
TC_{\alpha}\left({\bf x},{\bf w}\right)=\sum_{i=1}^D\left| \frac{x^{\alpha}_i}{\sum_{j=1}^Dx^{\alpha}_j}-
\frac{w^{\alpha}_i}{\sum_{j=1}^Dw^{\alpha}_j} \right|
\end{eqnarray}
\end{footnotesize}

\subsection{The Aitchisonian metric for compositional data}
\citet{ait2003} suggested the Euclidean metric applied to the log-ratio transformed data as a measure of distance between compositions
\begin{footnotesize}
\begin{eqnarray} \label{dist}
Ait\left({\bf x},{\bf w}\right)=\left[\sum_{i=1}^D\left(\log{\frac{x_i}{g\left({\bf x}\right)}}-\log{\frac{w_i}{g\left({\bf w}\right)}} \right)^2 \right]^{1/2},
\end{eqnarray} 
\end{footnotesize}
where $g\left({\bf z} \right)=\prod_{i=1}^Dz_i^{1/D}$ stands for the geometric mean.

\subsection{Some comments}
The power transformed compositional vectors still sum to $1$ and thus the ES-OV$_{\alpha}$ (\ref{ESOVa}) is still a metric. It becomes clear that when $\alpha=1$ we end up with the ES-OV metric (\ref{ESOV}). If on the other hand $\alpha=0$, then the distance is zero, since the compositional vectors become equal to the centre of the simplex. An advantage of the ES-OV$_{\alpha}$ metric (\ref{ESOVa}) over the Aitchisonian metric (\ref{dist}) is that the the first one is defined even when zero values are present. In this case the Aitchisonian metric (\ref{dist}) becomes degenerate and thus cannot be used. We have to note that we need to scale the data so that they sum to $1$ in the case of the ES-OV metric, but this is not a requirement of the taxicab metric.

Alternative metrics could be used as well, such as
\begin{enumerate}
\item the Hellinger metric \citep{owen2001}
\begin{footnotesize}
\begin{eqnarray*}
H\left({\bf x},{\bf w}\right)=\frac{1}{\sqrt{2}}\left[\sum_{i=1}^D\left(\sqrt{x_i}-\sqrt{w_i}\right)^2\right]^{1/2}
\end{eqnarray*}
\end{footnotesize}
\item or the angular metric if we treat compositional data as directional data (for more information about this approach see \citet{stephens1982} and \citet{scealy2011b, scealy2012})
\begin{footnotesize}
\begin{eqnarray*} 
Ang\left({\bf x},{\bf w}\right)=\arccos{\left(\sum_{i=1}^Dx_iw_i\right)}
\end{eqnarray*}
\end{footnotesize}
\end{enumerate} 

\citet{ait1992} argued that a simplicial metric should satisfy certain properties. These properties include 

\begin{enumerate}
\item \textit{Scale invariance}. The requirement here is that the measure used to define the distance between two compositional data vectors should be scale invariant, in the sense that it makes no difference whether the compositions are represented by proportions or percentages.
\item \textit{Subcompositional dominance}. To explain this we consider two compositional data vectors and we select sub-vectors from each consisting of the same components. Subcompositional dominance means that the distance between the sub-vectors is always less than or equal to the distance between the original compositional vectors.
\item \textit{Perturbation invariance}. The requirement here is that the distance between compositional vectors ${\bf x}$ and ${\bf w}$ should be the same as distance between ${\bf x}\oplus_0{\bf p}$ and ${\bf w}\oplus_0{\bf p}$, where the operator $\oplus_0$ means element-wise multiplication and then division by the sum so that the resulting vectors belong to $S^d$ and ${\bf p}$ is any vector (not necessarily compositional) with positive components.
\end{enumerate} 

If all of the above metrics satisfy or not these thee properties should not be a problem. Take for example subcompositional dominance. If someone has a compositional dataset, there has to be a good reason why he would choose to discard some components and form a sub-composition. And even if he does, all the metrics are still applicable. 

The message this paper tries to convey is that if someone uses a well defined metric (or even a dissimilarity measure) in order to perform classification he should be fine with that. When dealing with data lying on the Euclidean space, one can use dissimilarity measures as well to perform clustering or discrimination. The question of interest is how can we discriminate the observed groups of points as adequately as possible.  

\subsection{Loci of points equidistant from the centre of the simplex}
\begin{figure}[!ht]
\centering
\begin{tabular}{ccc}
\includegraphics[scale=0.28,trim=0 20 0 20]{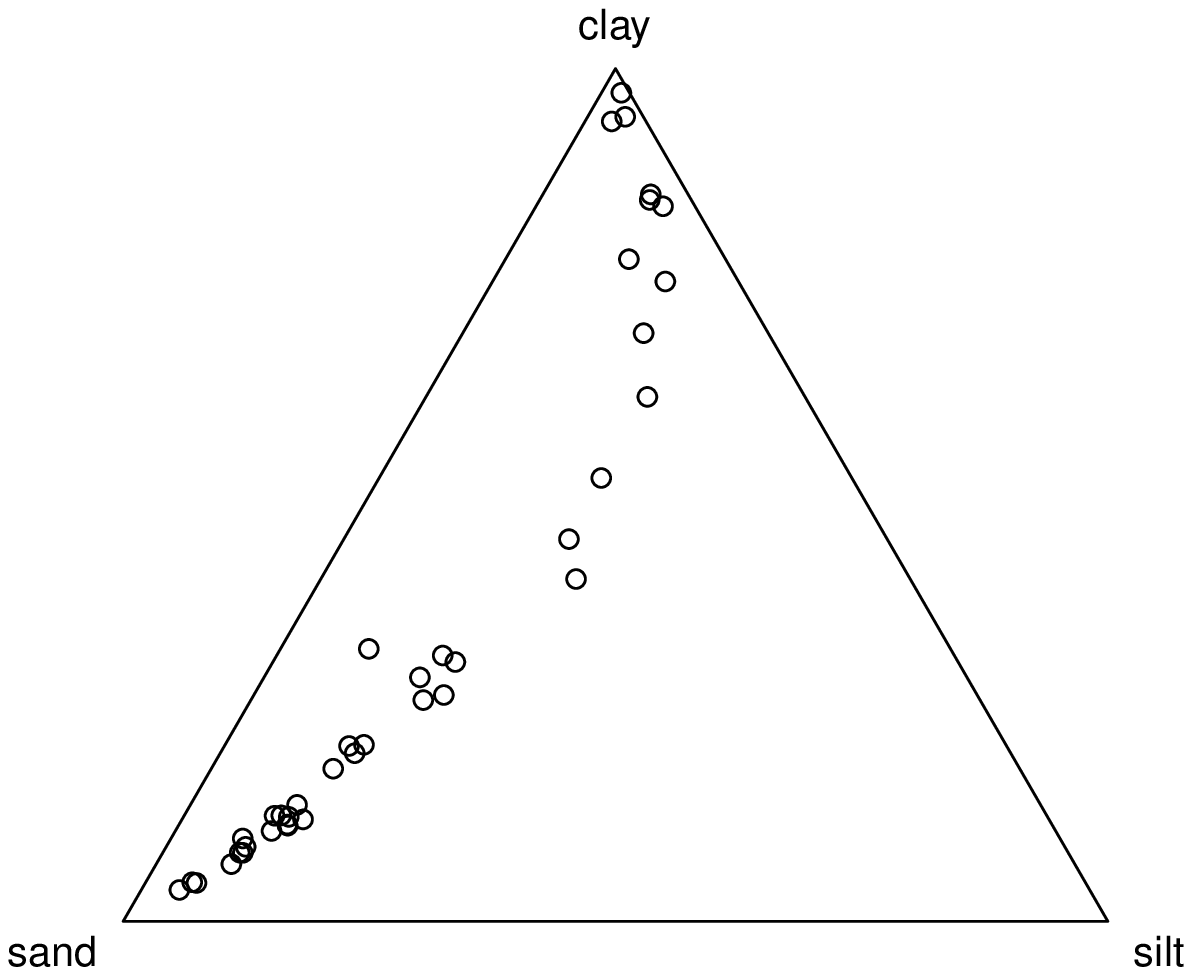} &
\includegraphics[scale=0.28,trim=0 20 0 20]{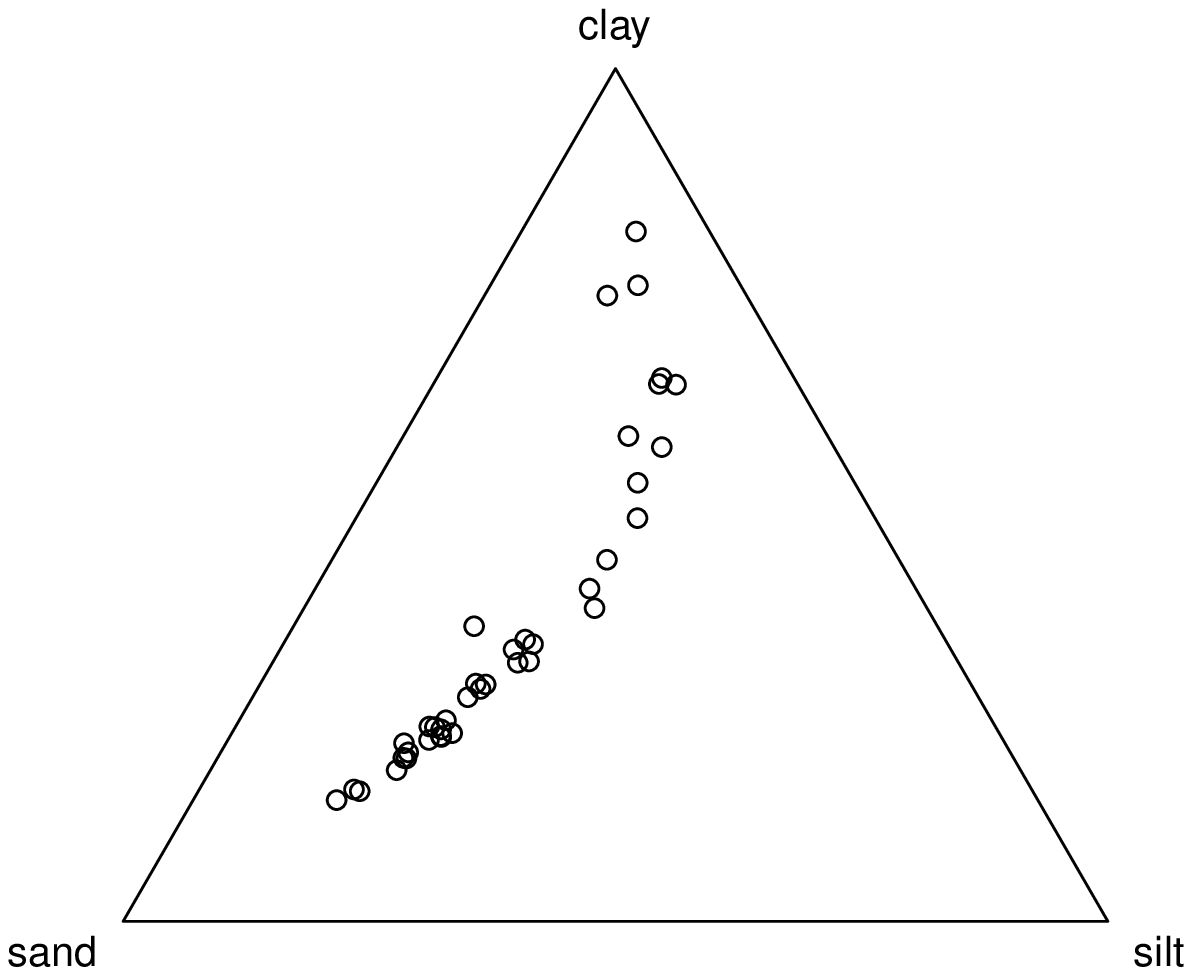} &
\includegraphics[scale=0.28,trim=0 20 0 20]{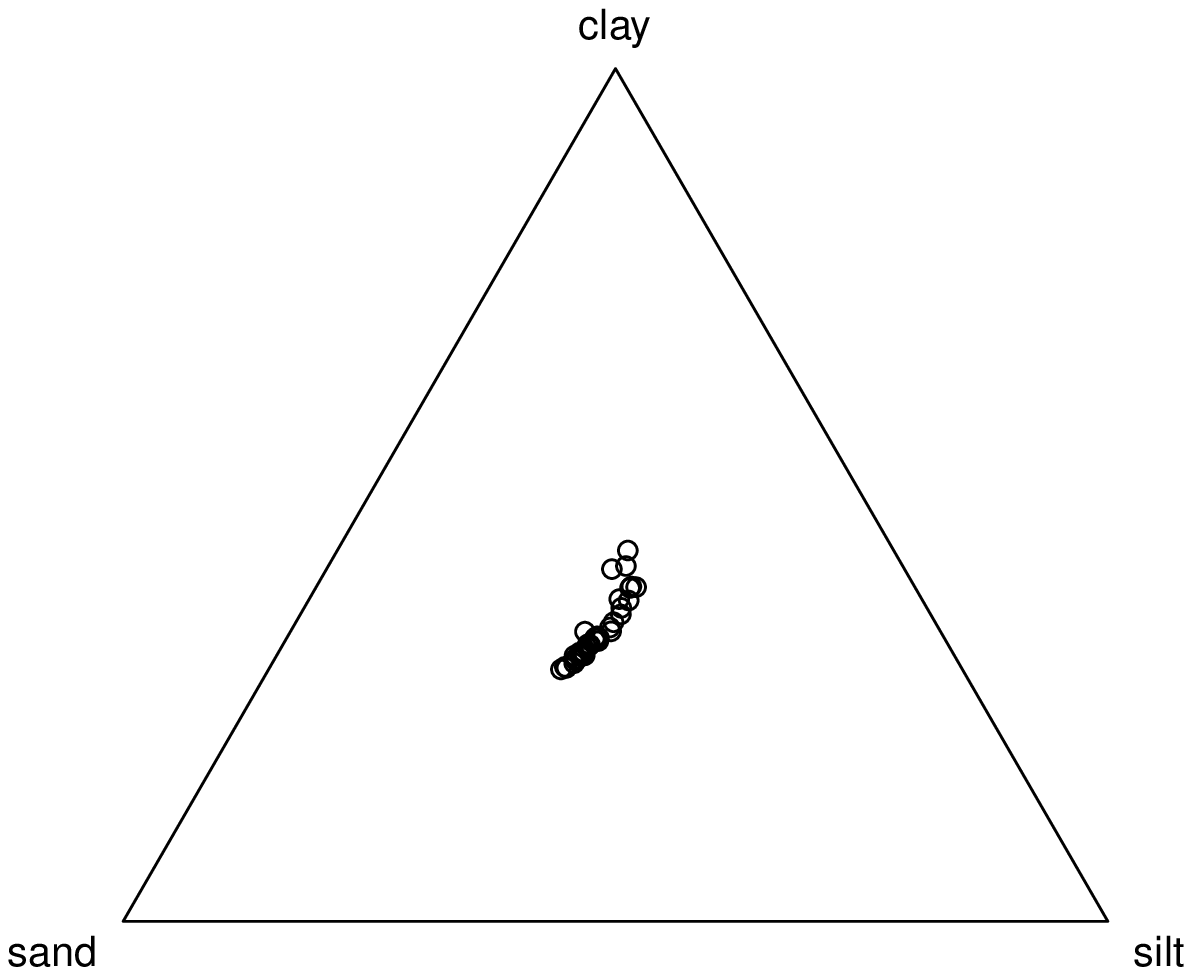}  \\
\footnotesize{(a)}   &  \footnotesize{(b)}   &  \footnotesize{(c)} \\
\includegraphics[scale=0.28,trim=0 20 0 20]{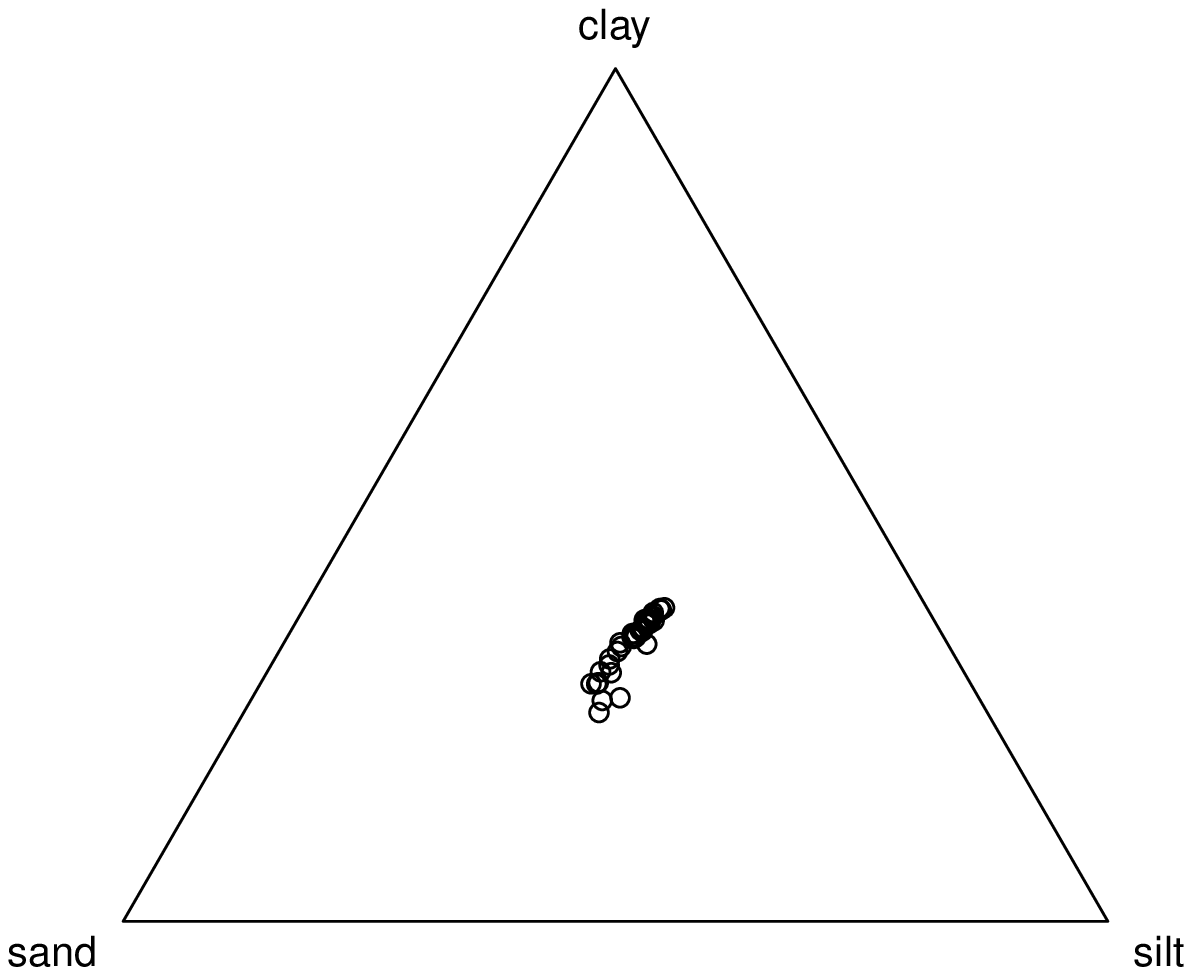} &
\includegraphics[scale=0.28,trim=0 20 0 20]{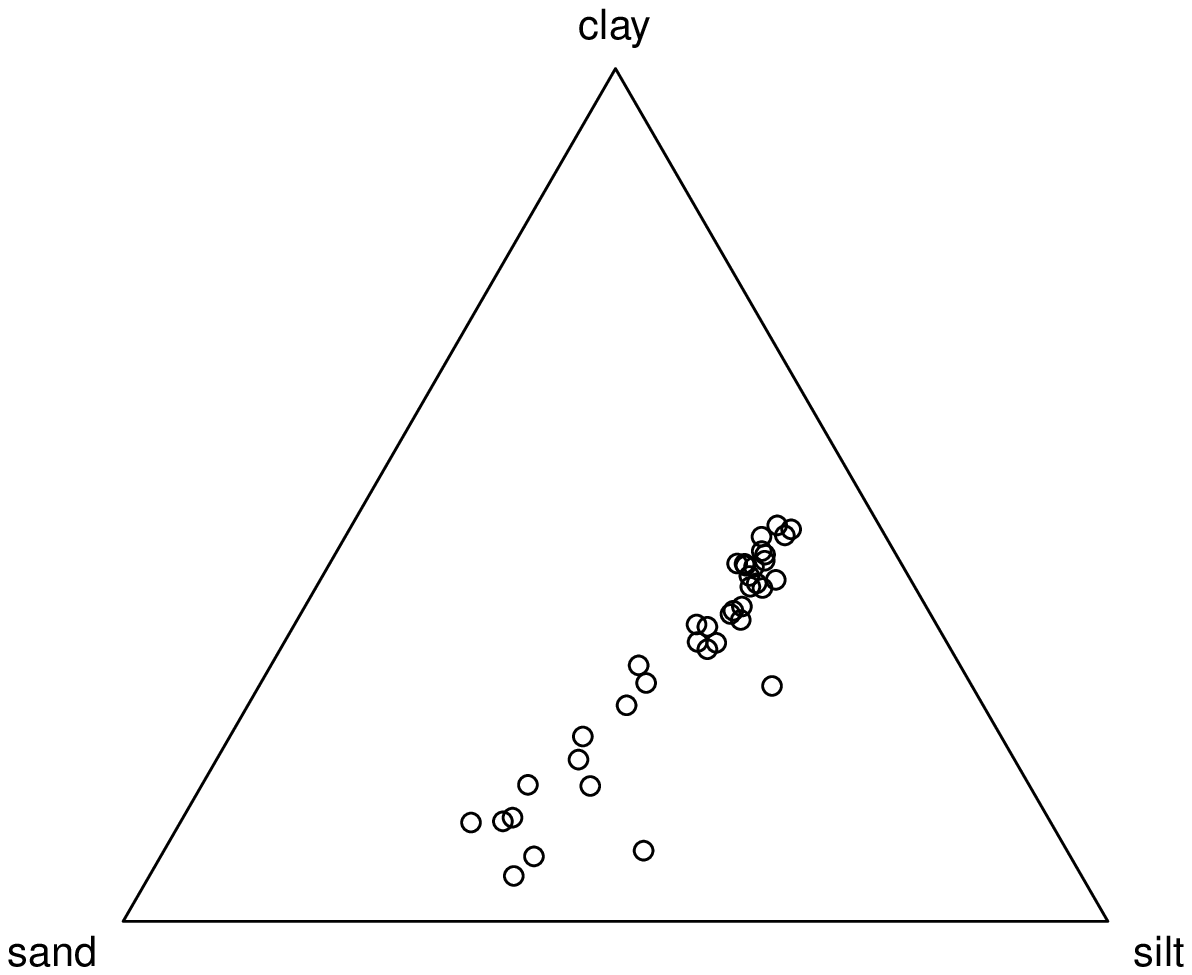} &
\includegraphics[scale=0.28,trim=0 20 0 20]{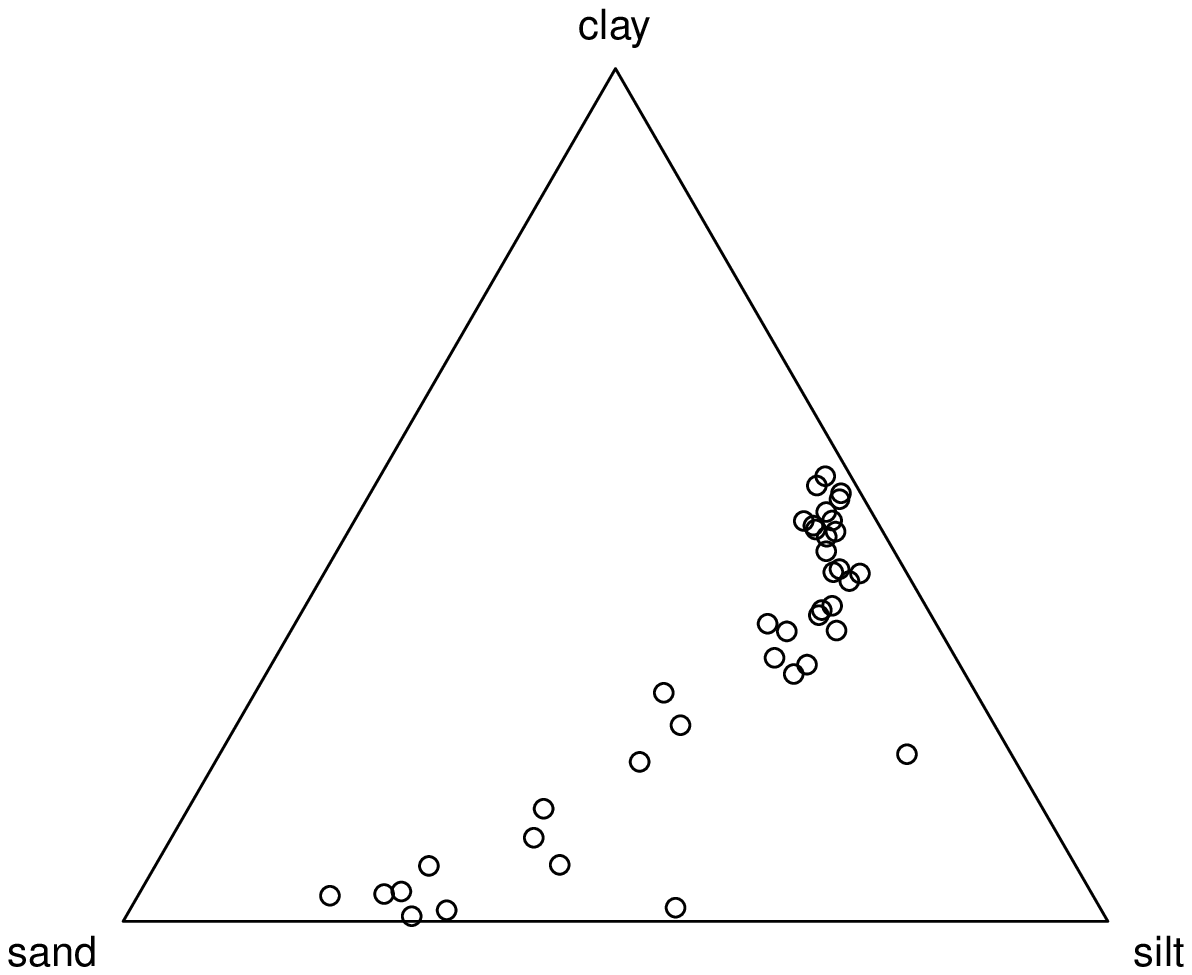}  \\
\footnotesize{(d)}   &  \footnotesize{(e)}   &  \footnotesize{(f)} 
\end{tabular}
\caption{Ternary plots of the Arctic lake data \citep{ait2003} for different values of $\alpha$. The data are transformed calculated using (a) $\alpha=-1$, (b) $\alpha=-0.5$, (c) $\alpha=-0.1$, (d) $\alpha=0.1$, (e) $\alpha=0.5$ and (f) $\alpha=1$.}
\label{fig1}
\end{figure}

\begin{figure}[!ht]
\centering
\begin{tabular}{ccc}
\includegraphics[scale=0.27,trim=0 20 0 20]{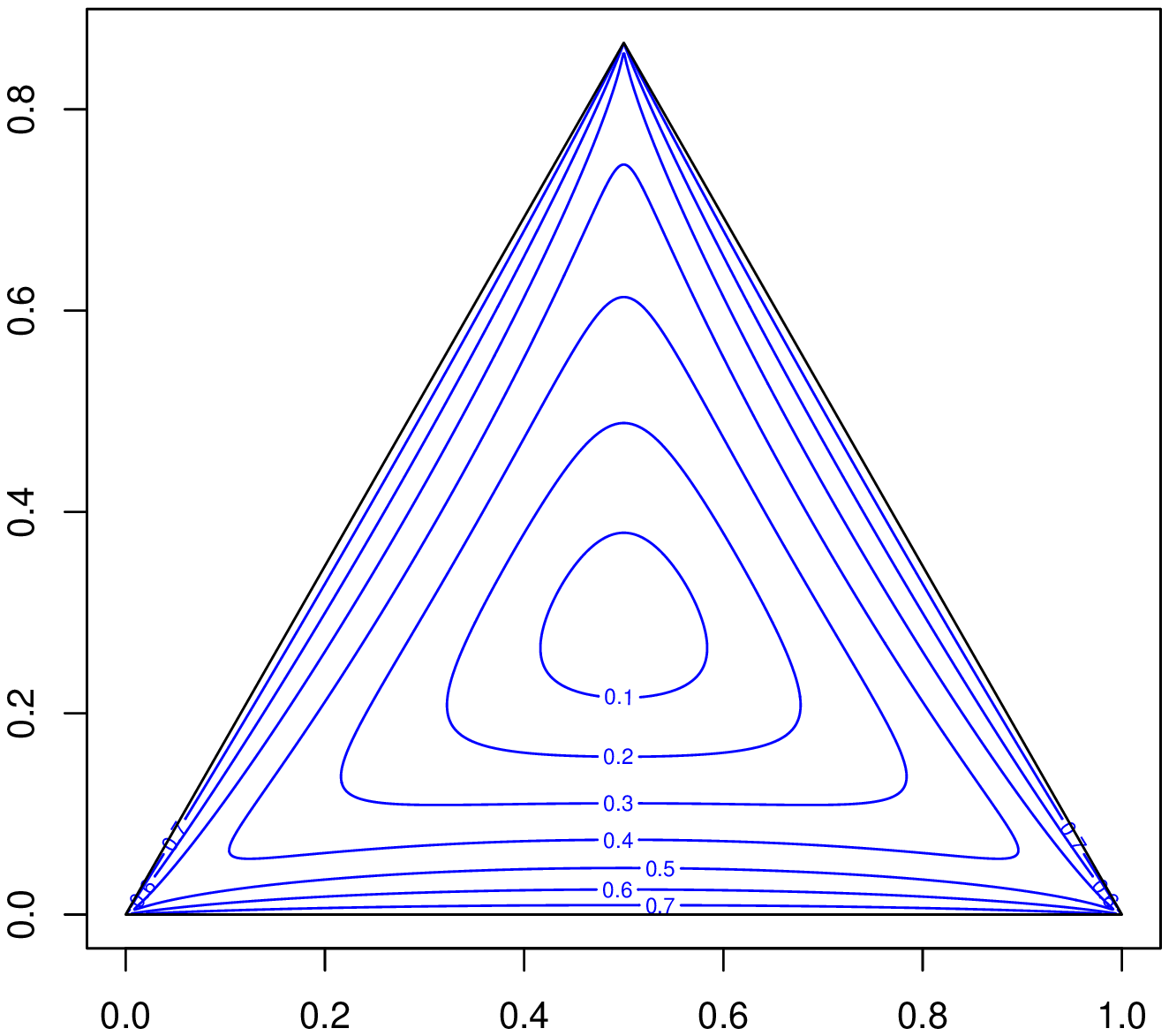} &
\includegraphics[scale=0.27,trim=0 20 0 20]{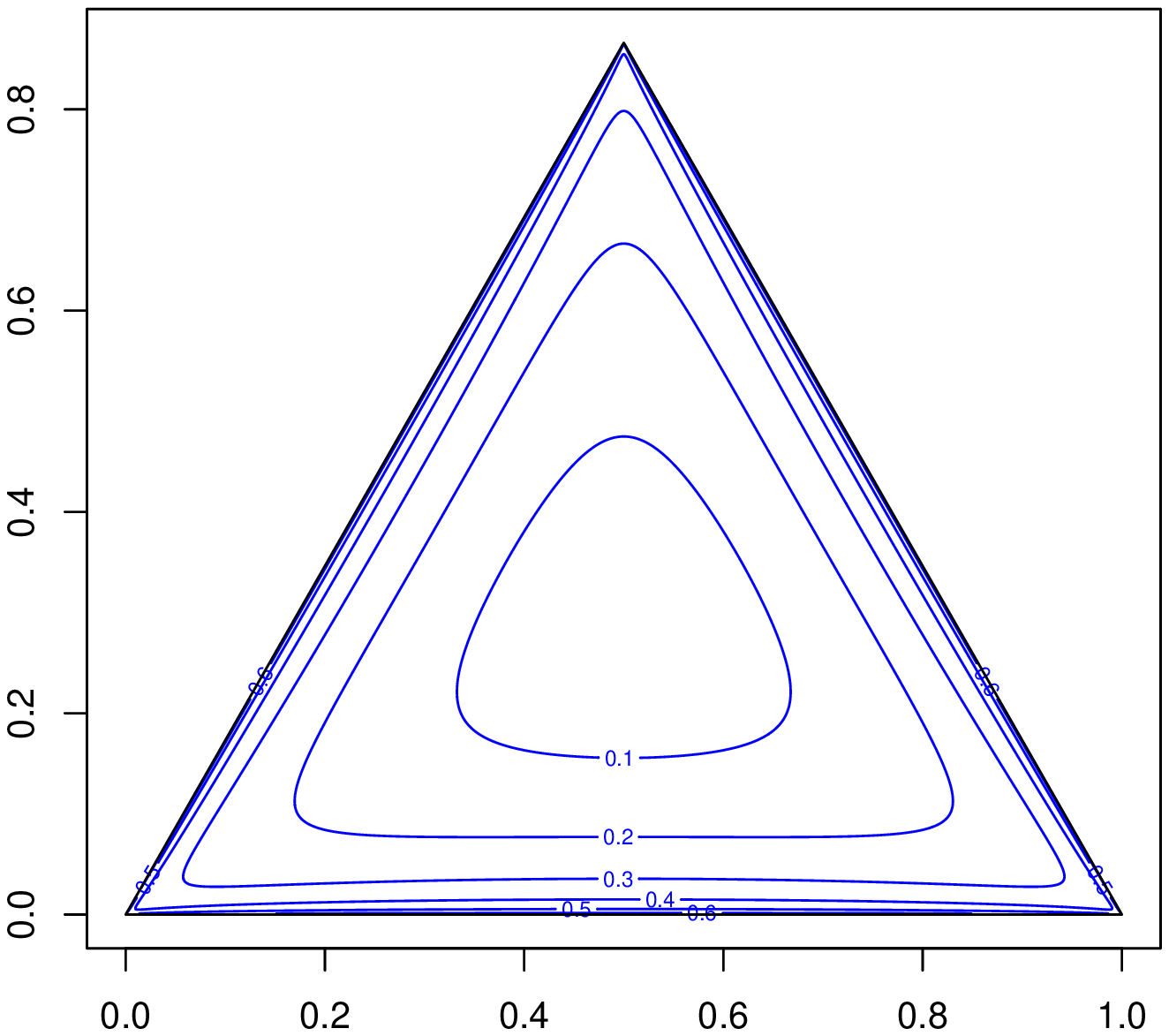} &
\includegraphics[scale=0.27,trim=0 20 0 20]{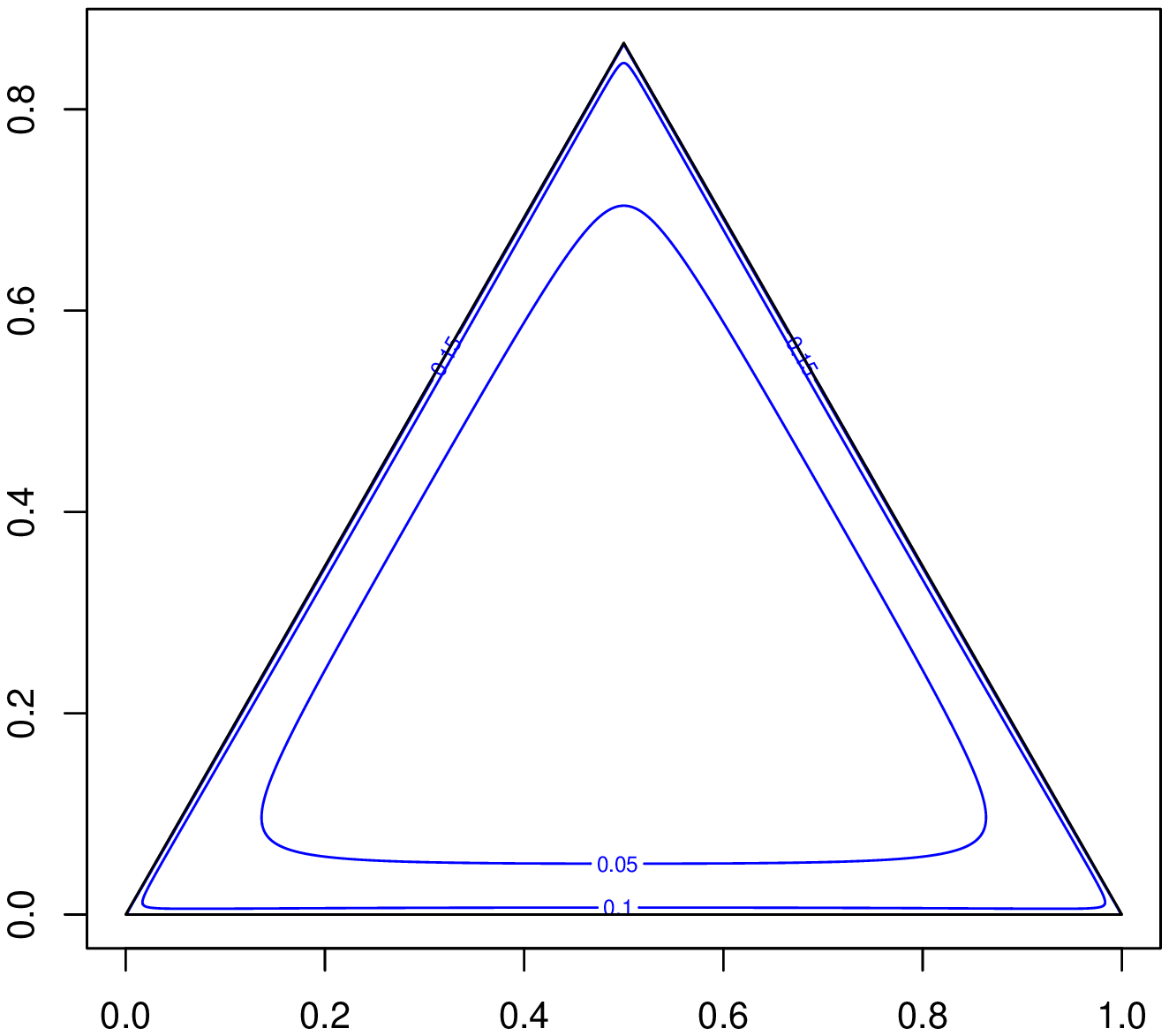}  \\
\footnotesize{(a)}   &  \footnotesize{(b)}   &  \footnotesize{(c)} \\
\includegraphics[scale=0.27,trim=0 20 0 20]{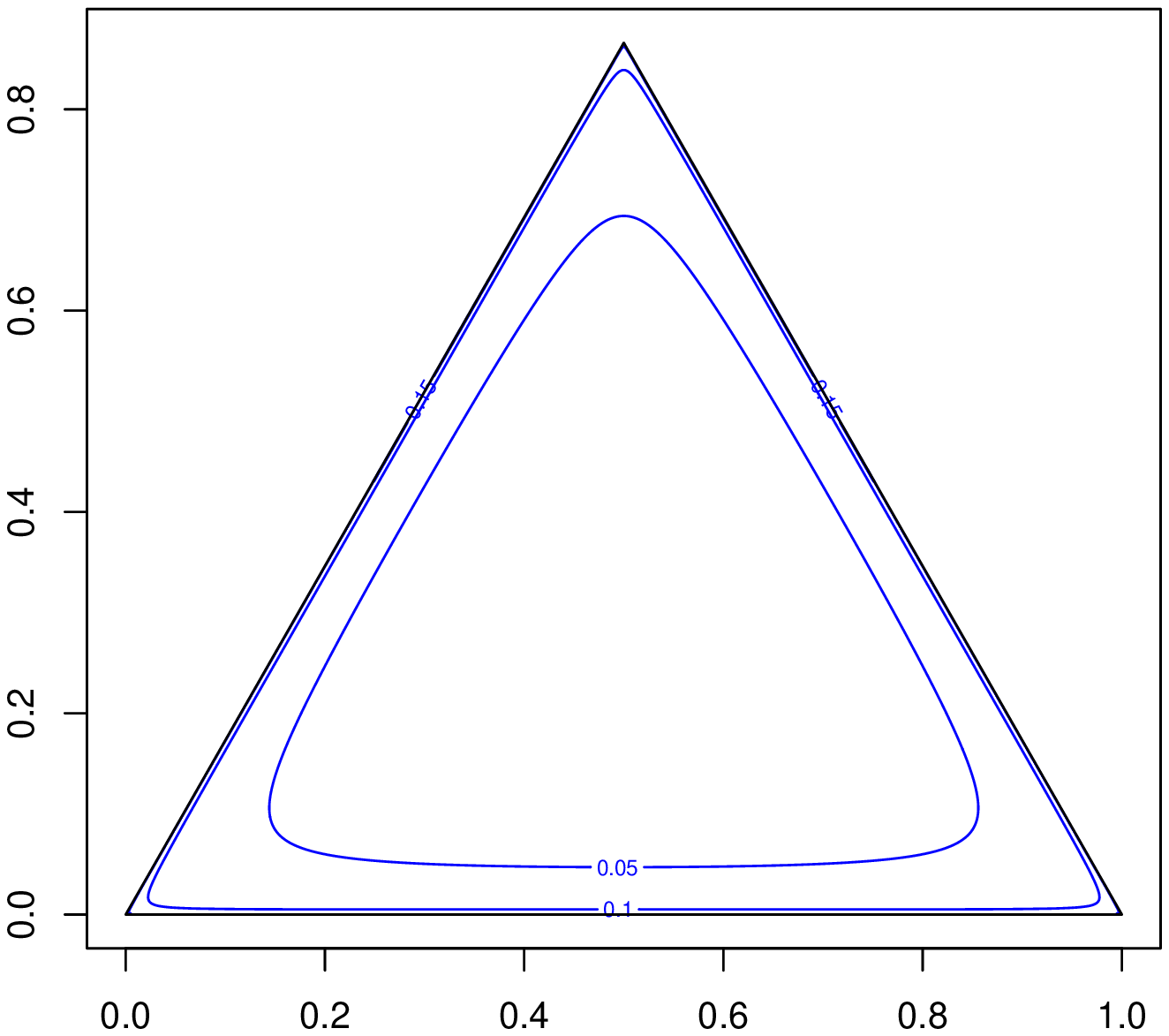} &
\includegraphics[scale=0.27,trim=0 20 0 20]{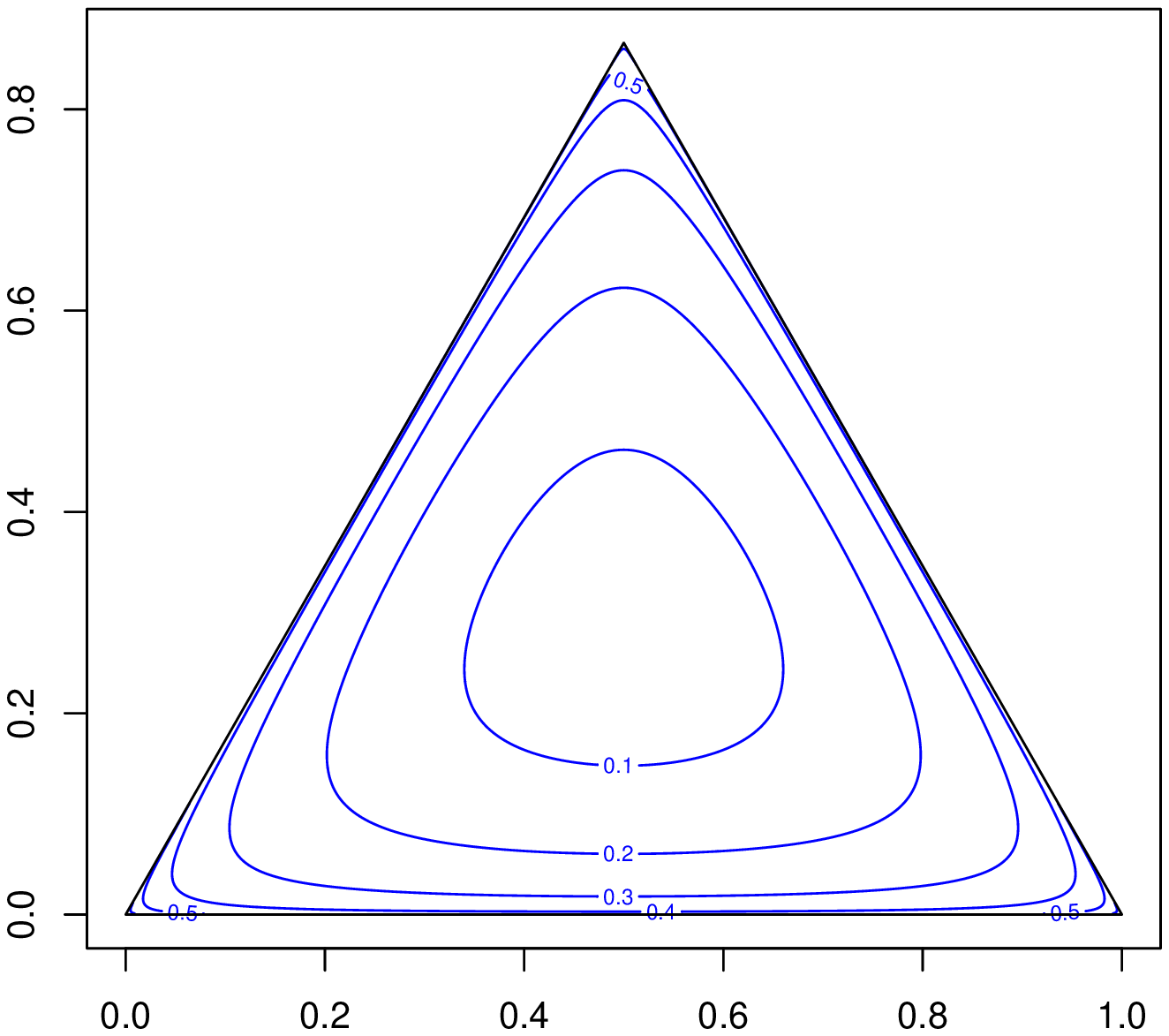} &
\includegraphics[scale=0.27,trim=0 20 0 20]{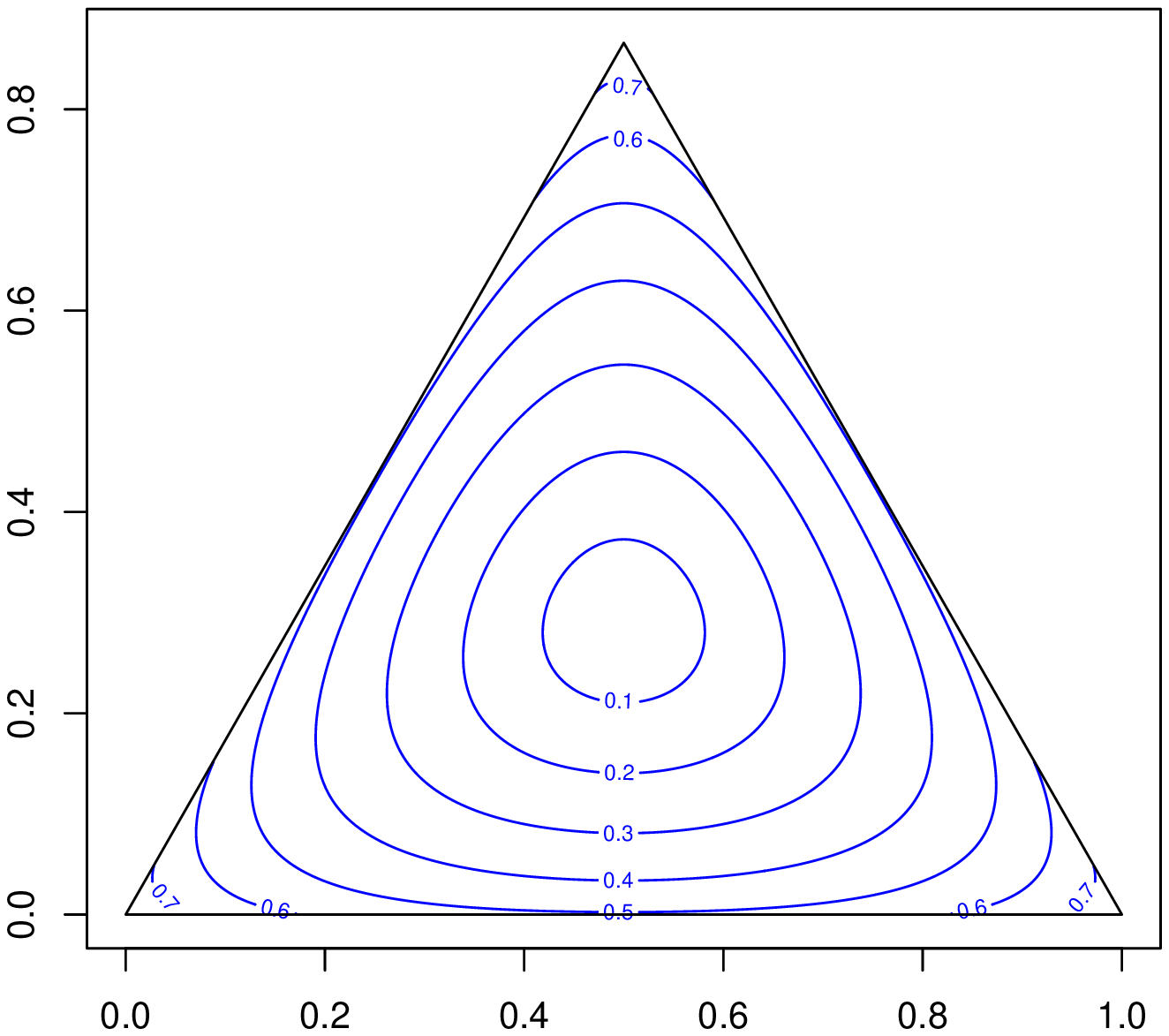}  \\
\footnotesize{(d)}   &  \footnotesize{(e)}   &  \footnotesize{(f)} 
\end{tabular}
\caption{Loci of points equidistant from the centre of the simplex using the ESOV$_{\alpha}$ metric (\ref{ESOVa}). In all cases the distances are from the barycentre of the simplex $\left(1/3,1/3,1/3\right)$. The contours are calculated using (a) $\alpha=-1$, (b) $\alpha=-0.5$, (c) $\alpha=-0.1$, (d) $\alpha=0.1$, (e) $\alpha=0.5$ and (f) $\alpha=1$.}
\label{fig2}
\end{figure}

\begin{figure}[!ht]
\centering
\begin{tabular}{ccc}
\includegraphics[scale=0.27,trim=0 20 0 20]{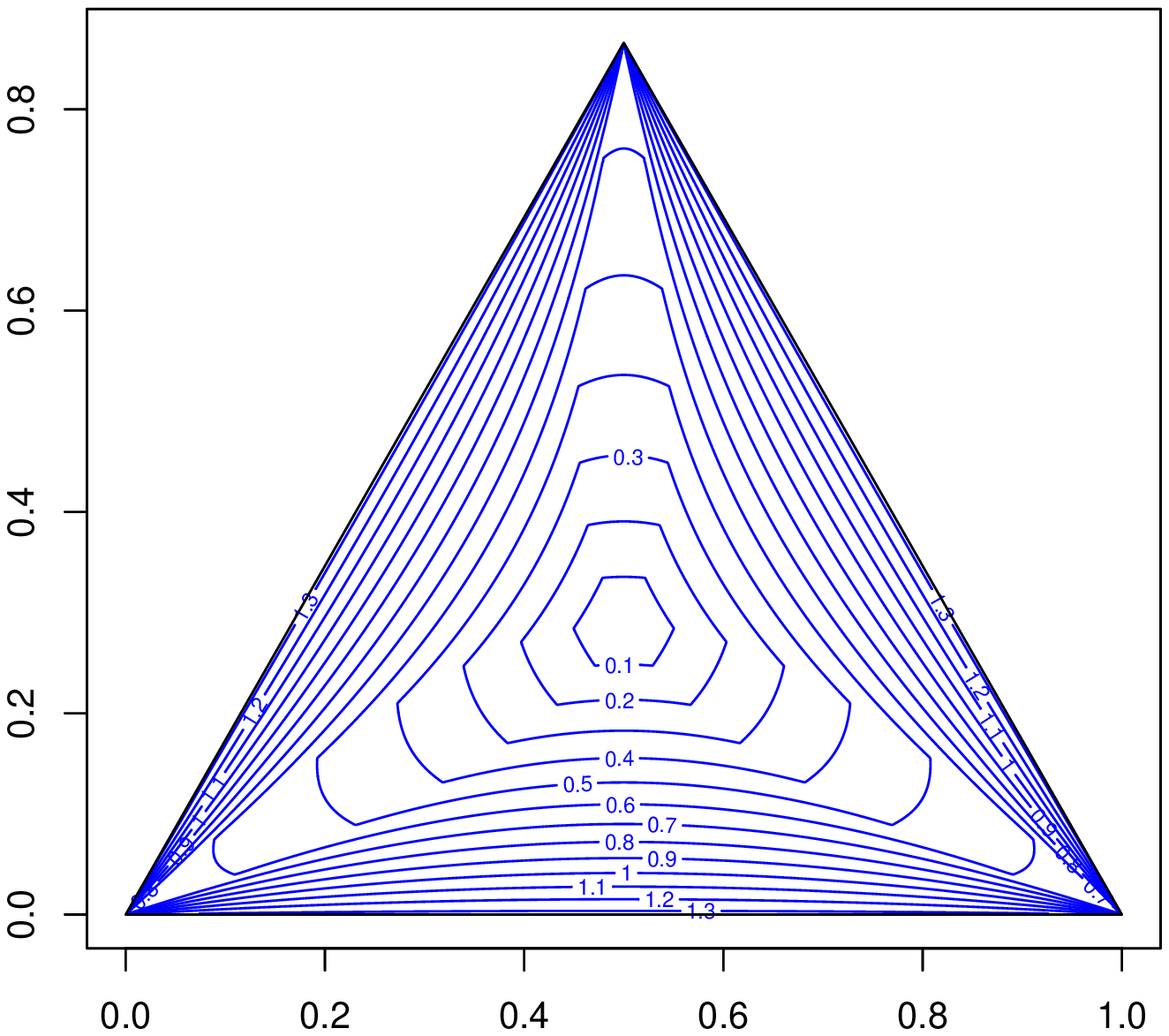} &
\includegraphics[scale=0.27,trim=0 20 0 20]{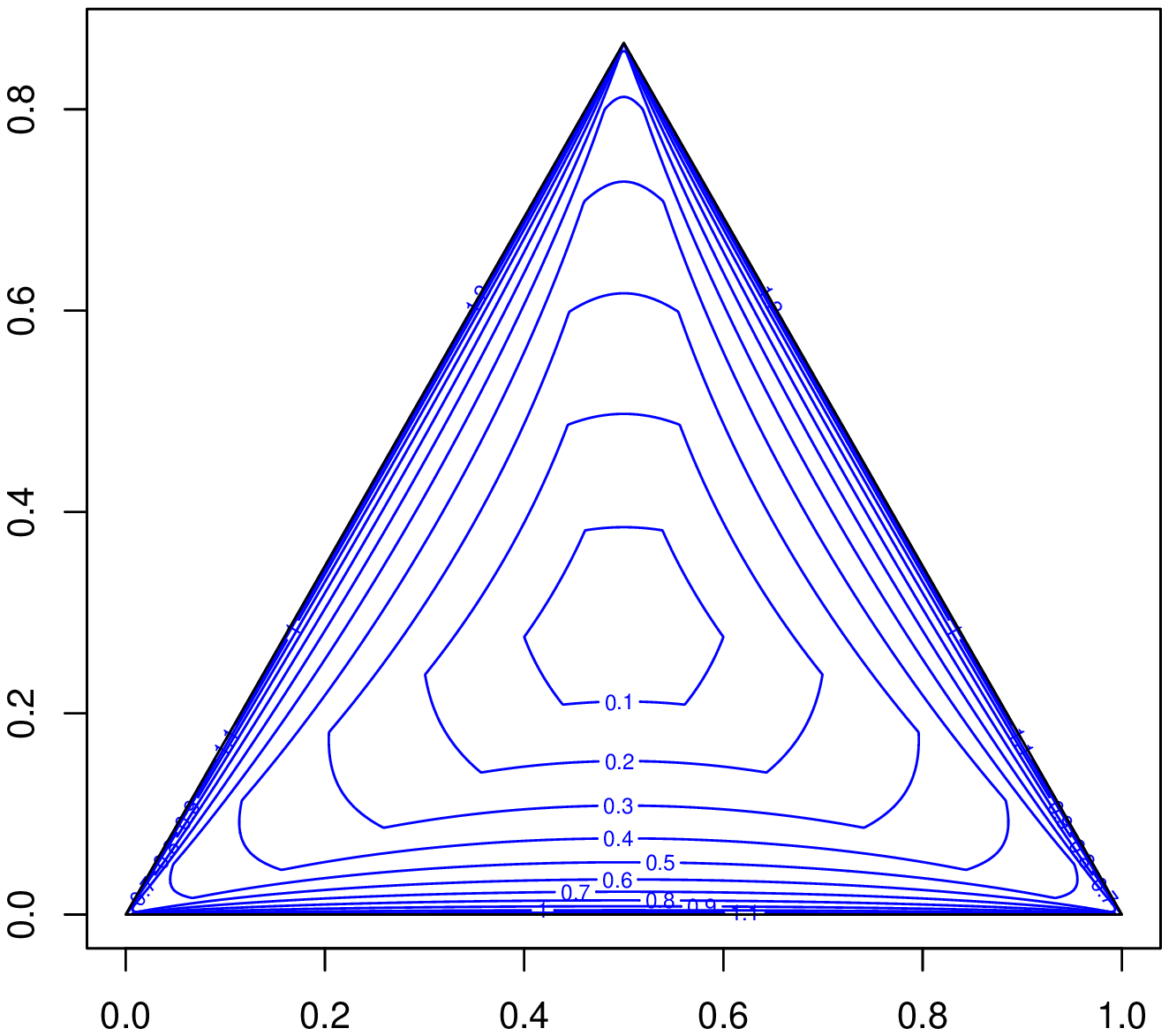} &
\includegraphics[scale=0.27,trim=0 20 0 20]{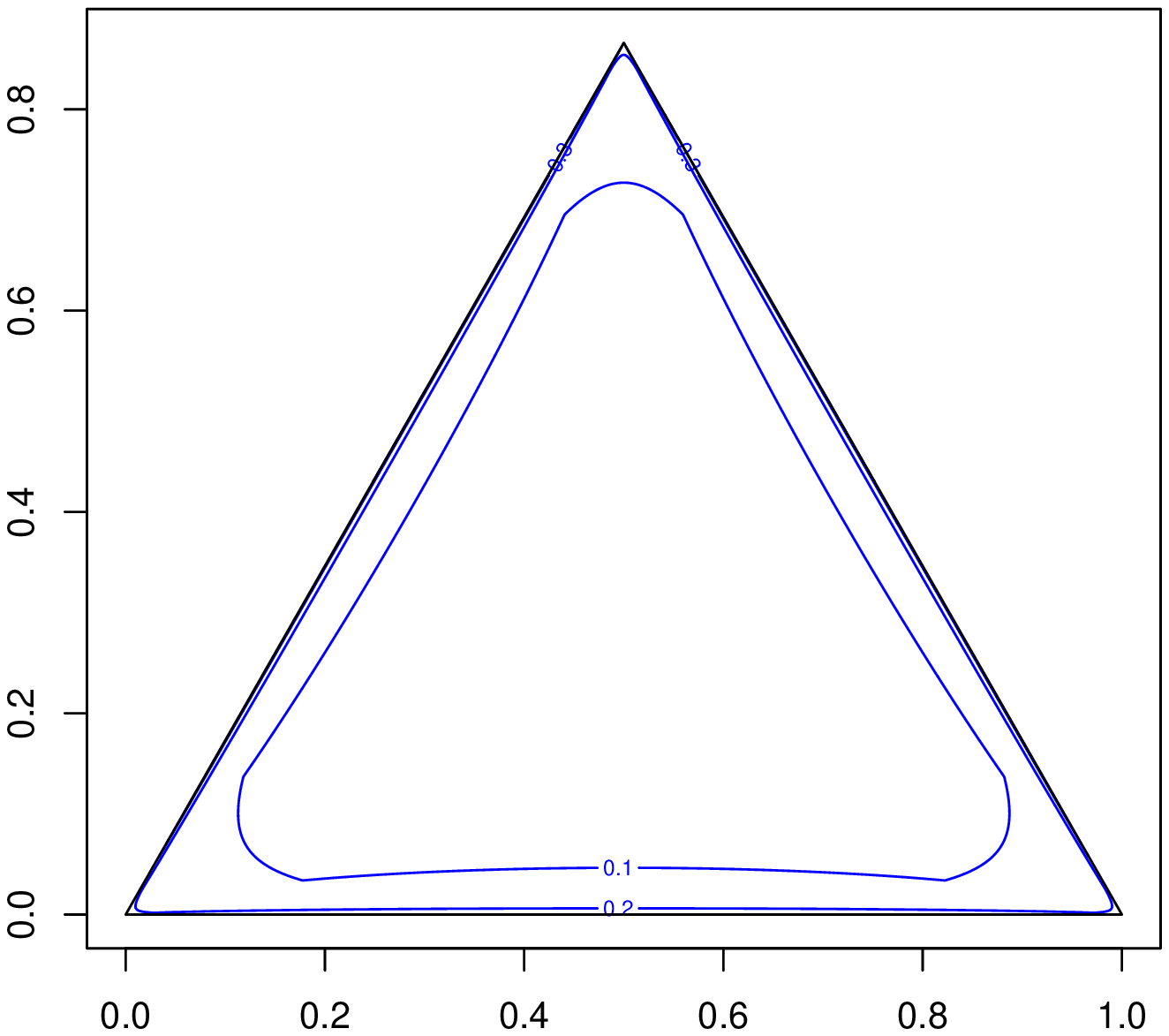}  \\
\footnotesize{(a)}   &  \footnotesize{(b)}   &  \footnotesize{(c)} \\
\includegraphics[scale=0.27,trim=0 20 0 20]{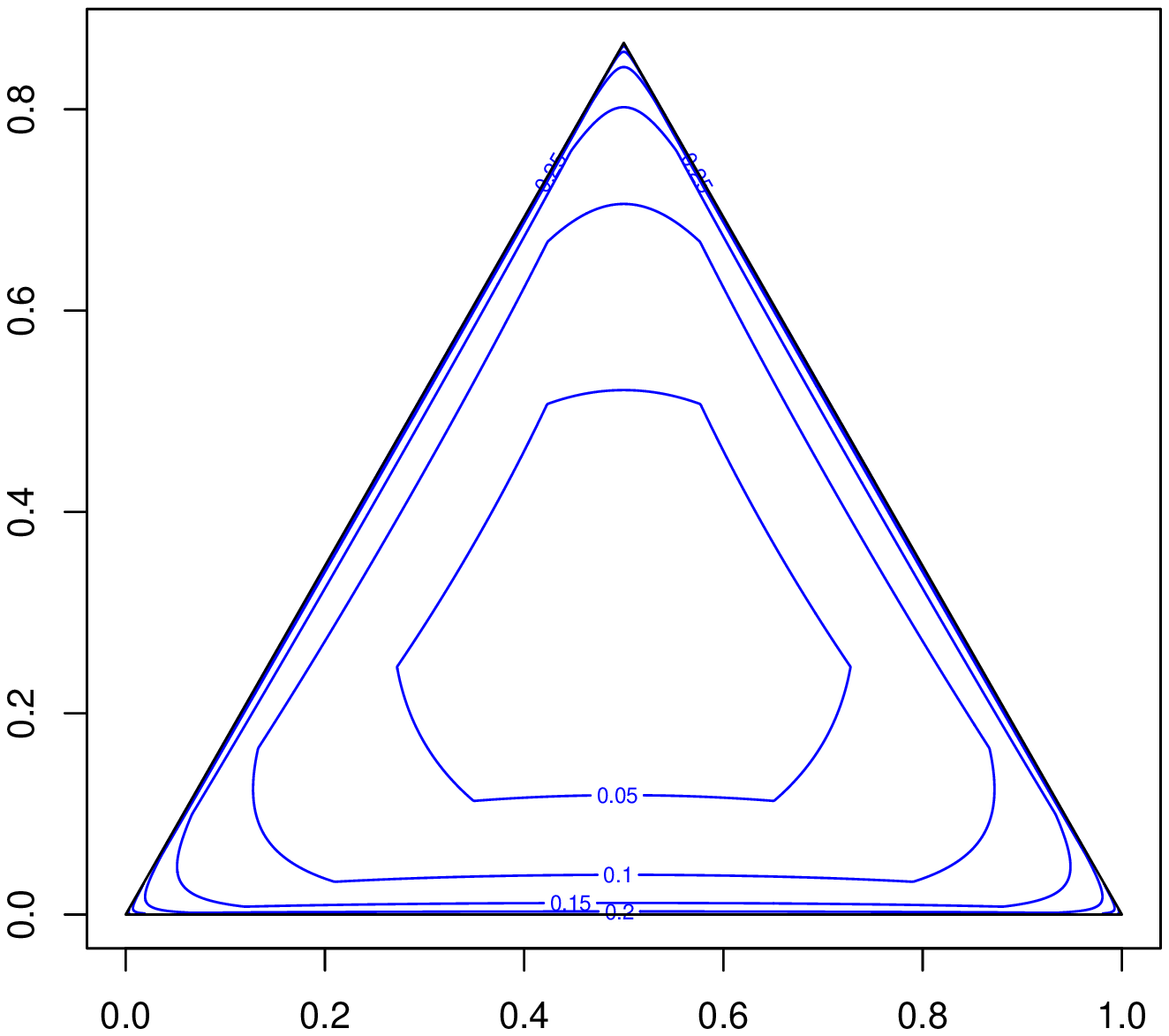} &
\includegraphics[scale=0.27,trim=0 20 0 20]{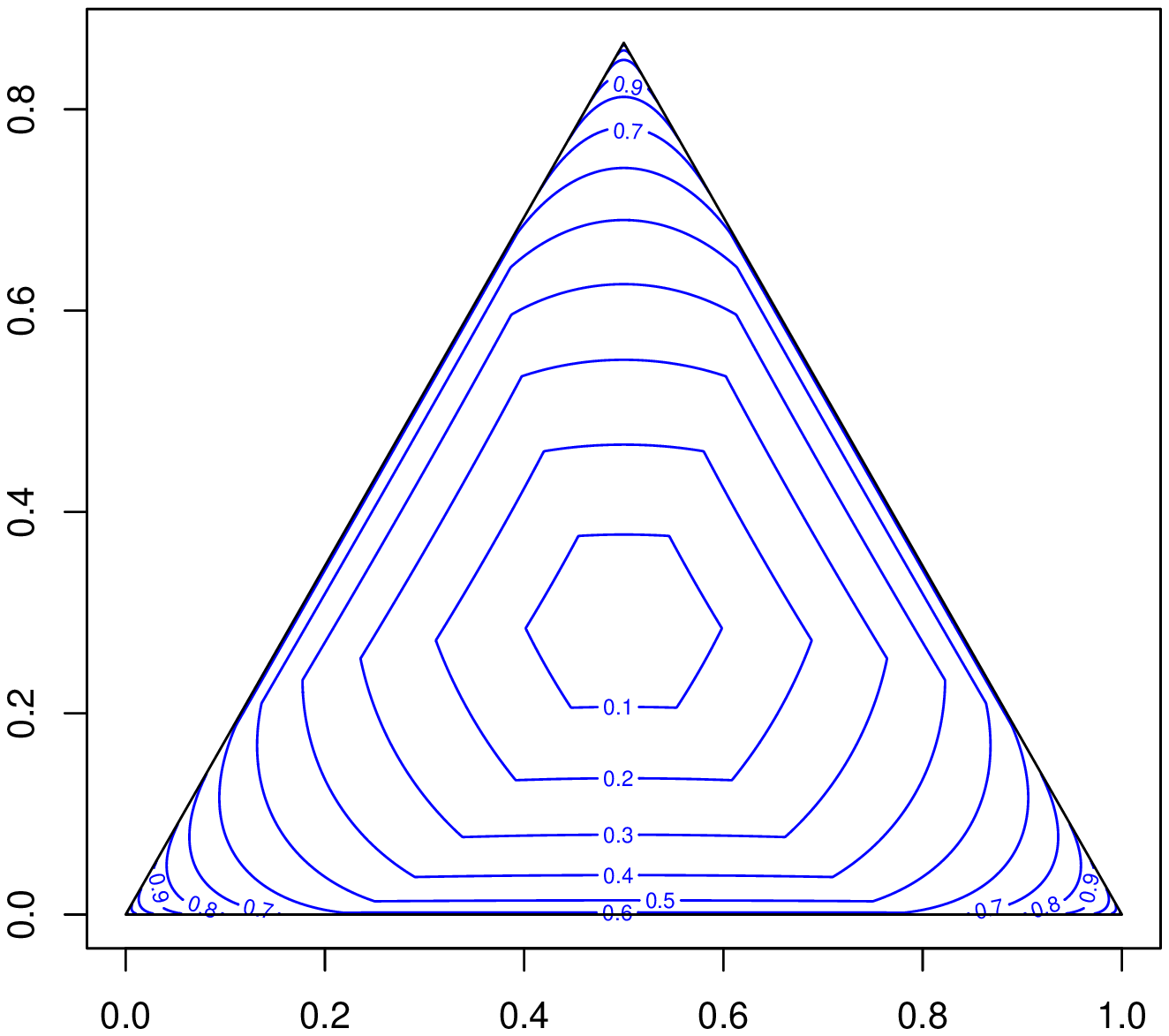} &
\includegraphics[scale=0.27,trim=0 20 0 20]{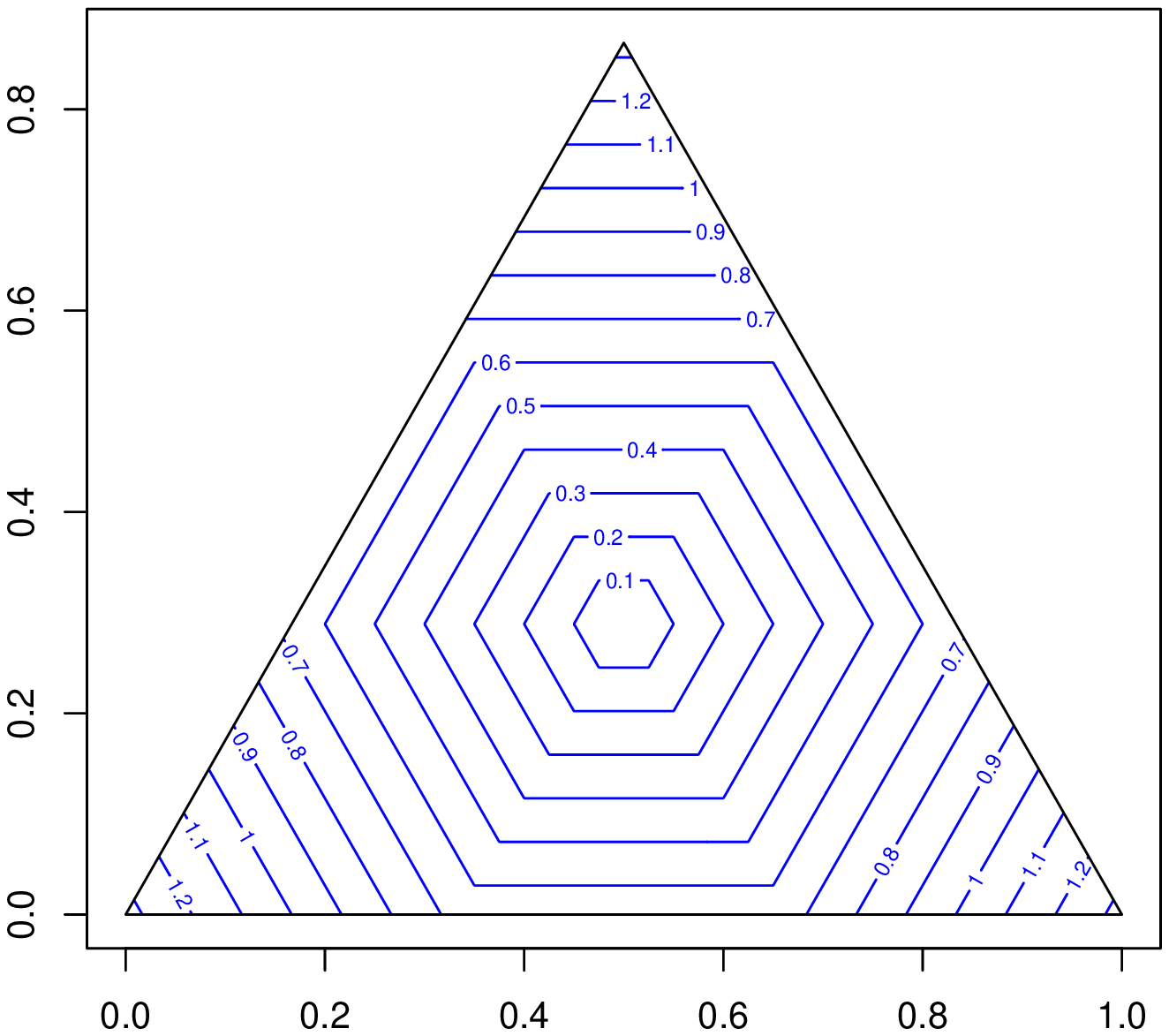}  \\
\footnotesize{(d)}   &  \footnotesize{(e)}   &  \footnotesize{(f)} \\
\end{tabular}
\caption{Loci of points equidistant from the centre of the simplex using the TC$_{\alpha}$ metric (\ref{taxicaba}). In all cases the distances are from the barycentre of the simplex $\left(1/3,1/3,1/3\right)$. The contours are calculated using (a) $\alpha=-1$, (b) $\alpha=-0.5$, (c) $\alpha=-0.1$, (d) $\alpha=0.1$, (e) $\alpha=0.5$ and (f) $\alpha=1$.}
\label{fig3}
\end{figure}

\begin{figure}[!ht]
\centering
\includegraphics[scale=0.35,trim=0 20 0 20]{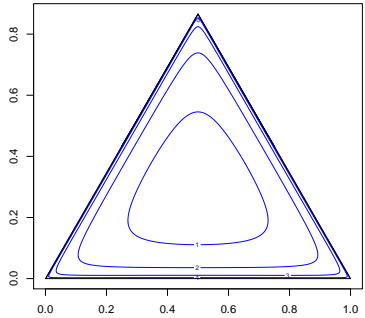}
\caption{Loci of points equidistant from the centre of the simplex using the Aitchisonian metric (\ref{dist}).} 
\label{fig4}
\end{figure}

Figure \ref{fig1} shows the effect of the power transformation (\ref{alpha}) on the data. As expected, the data come closer to the barycentre of the triangle as $\alpha$ tends to zero. The data used and plotted on Figure \ref{fig1} are the Arctic lake data \citep{ait2003}. Figures \ref{fig2} and \ref{fig3} show the plots of loci of points of the ES-OV$_{\alpha}$ metric (\ref{ESOVa}) and of the TC$_{\alpha}$ metric (\ref{taxicaba}) for different values of $\alpha$ and Figure \ref{fig4} shows the contour plots of the Aitchisonian metric (\ref{dist}). In all cases, the plots of loci of points refer to the distance from the barycentre of the simplex. The loci of points seen on Figure \ref{fig2} have similar shape regardless of the value of $\alpha$. This is not true for the loci in Figure \ref{fig3}, which change as the value of $\alpha$ changes. 

\section{Supervised classification for compositional data using the $k$-NN algorithm}
The goal of this paper is to perform supervised classification of compositional data using the $k$-NN algorithm. For this reason we will use the ES-OV$_{\alpha}$ (\ref{ESOVa}) and TC$_{\alpha}$ (\ref{taxicaba}) metrics and compare their performance and suitability with the Aitchisonian metric metric (\ref{dist}). 

The $k$-NN algorithm is a non-parametric supervised learning technique which is computationally heavier than quadratic and linear  discriminant analysis but easier to implement as it relies solely on metrics between points. 

Similarly to other supervised classification techniques it requires some parameter tuning. The two parameters associated with it in our case are the power parameter $\alpha$ and the number of nearest neighbours $k$. We describe the steps of the $k$-NN for compositional data in our case. 

\begin{enumerate}
\item Separate the data into the training and the test dataset. 
\item Choose a value of $k$, the number of nearest neighbours.
\item Classify the test data using either the ES-OV$_{\alpha}$ (\ref{ESOVa}), the TC$_{\alpha}$ (\ref{taxicaba}) for a range of values of $\alpha$ and each time calculate the percentage of correct classification.
\item Repeat steps $2-3$ for a different value of $k$.
\item Repeat steps $1-4$ B (in our case $B=200$) times and for each $\alpha$ and $k$ and estimate the percentage of correct classification by averaging over all B times. 
\end{enumerate}

We can of course use the Aitchisonian metric (\ref{dist}) instead of the ES-OV$_\alpha$ (\ref{ESOVa}) or the TC$_{\alpha}$ metric (\ref{taxicaba}). In this case we have to choose the number of nearest neighbours only, since no power transformation is involved. We could of course use any other metric defined in $R^d$. In this case we would have to apply the additive log-ratio transformation \citep{ait2003} to the data. The issue in that case though would be the presence of zeros in the data. 

In the next section we will see two examples using real data and see the performance of the algorithm when each of the two metrics is used. 

\subsection{Examples using real data}
We will now see the performance of the $k$-NN algorithm using the ES-OV$_\alpha$ metric (\ref{ESOVa}), the TC$_{\alpha}$ metric and the Aitchisonian metric (\ref{dist}) with real data. 

\subsubsection*{\textit{Example 1. Hydrochemical data with no zero values}}
The first dataset comes from hydrochemistry. A hydrochemical data set \citep{otero2005} contains measurements on $14$ elements. the data were gathered within a period of $2$ years from $31$ stations located along the rivers and main tributaries of the Llobregat river, one of the medium rivers in northeastern Spain. Each of these elements is measured approximately once each month during these $2$ years. There are $4$ tributaries of interest, Anoia ($143$ measurements), Cardener ($95$ measurements), Upper Llobregat ($135$ measurements) and Lower Llobregat ($112$ measurements). Thus, there are $485$ across all $4$ tributaries. 

\begin{figure}[!ht]
\centering
\begin{tabular}{cc}
\includegraphics[scale=0.45,trim=0 20 0 20]{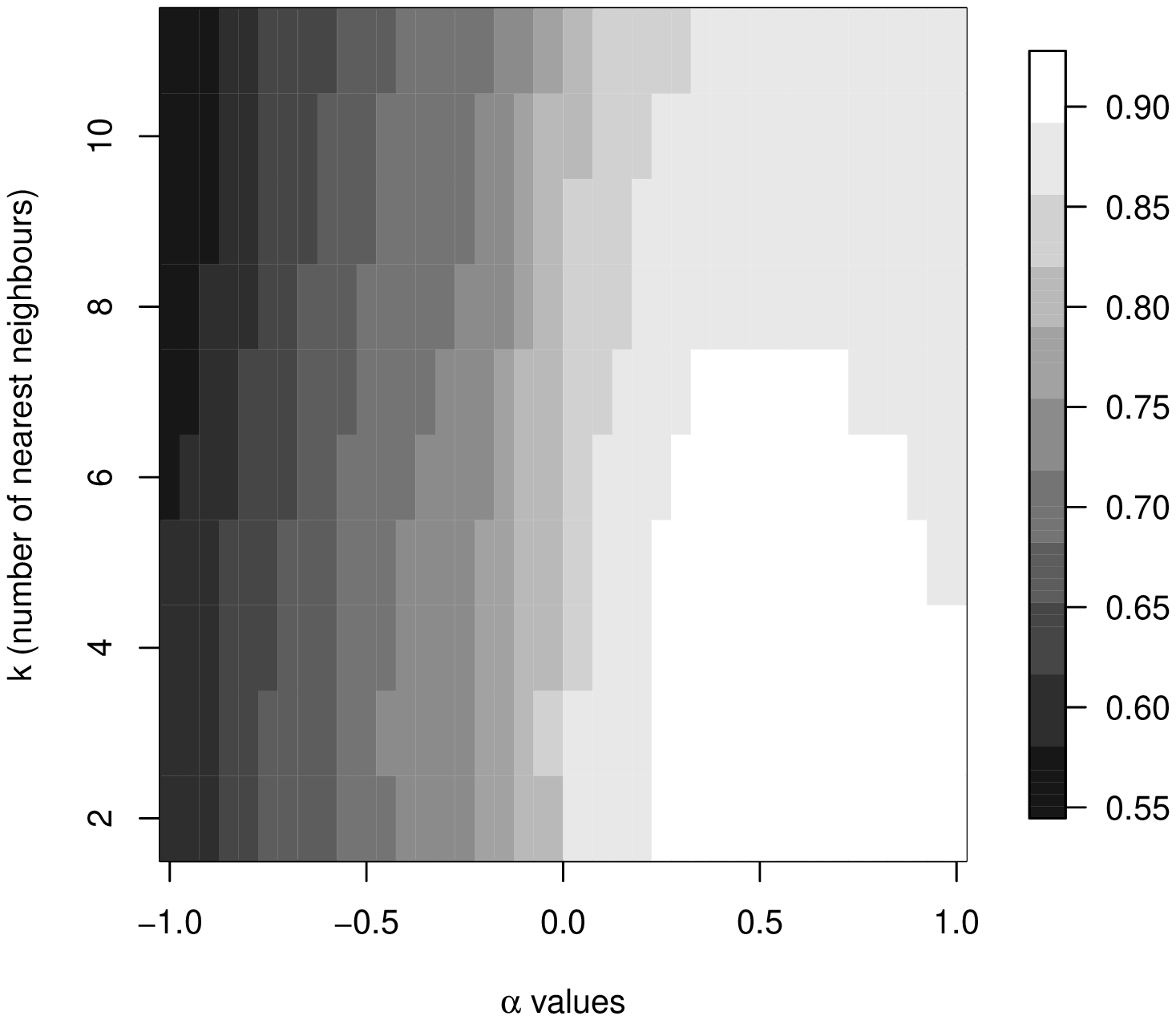} &
\includegraphics[scale=0.45,trim=0 20 0 20]{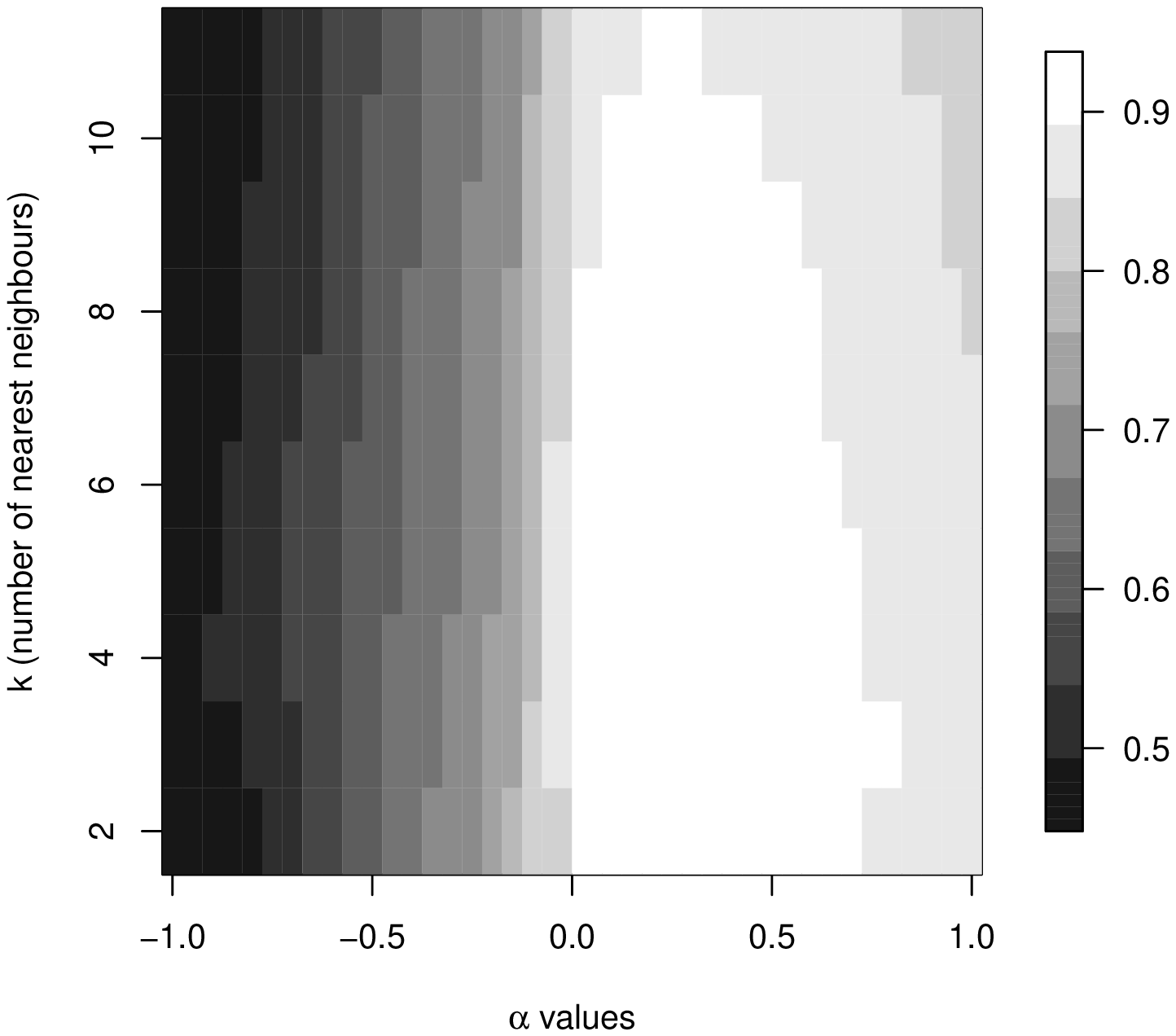}  \\
\footnotesize{(a)}   &  \footnotesize{(b)}  \\
\end{tabular}
\caption{The estimated percentage of correct classification for the hydrochemical data as a function of $k$, the nearest neighbours and of $\alpha$ using the (a) ES-OV$_{\alpha}$ metric (\ref{ESOVa}) and (b) TC$_{\alpha}$ (\ref{taxicaba}).} 
\label{fig5}
\end{figure}

This dataset contains no zero values, so all three metrics are applicable. The size of the training sample was equal to $434$ and thus the test sample consisted of $51$ observations, which were sampled using stratified random sampling each time to ensure that observations from all tributaries are selected every time. Figure \ref{fig5} shows the heat plot of the estimated percentage for different values of $k$ and $\alpha$. 

\begin{figure}[!ht]
\centering
\includegraphics[scale=0.45,trim=0 20 0 20]{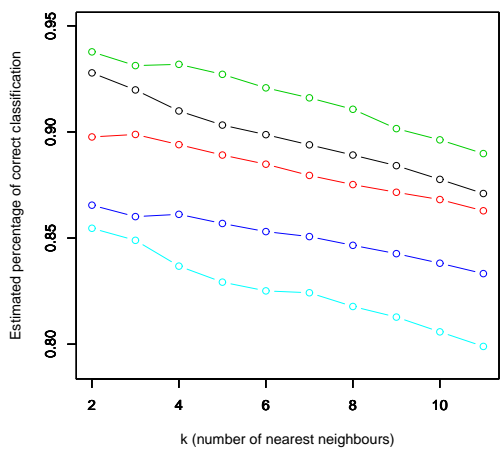}
\caption{The estimated percentage of correct classification as a function of $k$. The black and the red lines are based on the ES-OV$_{\alpha}$  metric (\ref{ESOVa}) with $\alpha=0.5$ and $\alpha=1$ respectively. The green and the blue lines are based on the TC$_{\alpha}$ metric (\ref{taxicaba}) with $\alpha=0.35$ and $\alpha=1$ respectively. The turquoise line is the Aitchisonian metric (\ref{dist}).} 
\label{fig6}
\end{figure}

If $\alpha=0.5$ and $k=2$ the estimated percentage percentage of correct classification is equal to $92.78\%$ and when $\alpha=1$ and $k=3$ the estimated percentage is $89.88\%$ when the ES-OV$_{\alpha}$ metric (\ref{ESOVa}) was applied. When the TC$_{\alpha}$ metric (\ref{taxicaba}) is applied the results are similar, with $\alpha=0.35$ and $k=2$ the estimated percentage of correct classification is $93.77\%$ and when $\alpha=1$ and $k=2$, the estimated percentage of correct classification is $86.55\%$. This is an example where a value of $\alpha$ other than $1$ leads to better results. The change in the percentage might seem small, but if we take into account the total sample size, we will see that the $3\%$ of $485$ observations is $14$ observations and it is not a small number. The Aitchisonian metric on the other hand did not do that well. The maximum estimated percentage was equal to $85.46\%$ when $k=2$. 

More information (including the specificities and sensitivities for each tributary averaged over all $200$ replications) regarding the classification results is presented in Table \ref{tab1} below. A general conclusion about the mean sensitivities and specificities is that the lower sensitivities are observed when the estimated percentage of correct classification is lower and they have also larger standard errors. The mean specificities on the other hand are in general high and are less affected by the estimated percentage of correct classification.

\begin{table}[h]
\begin{small}
\begin{center}
\begin{tabular}{ccccc} \hline
\multicolumn{5}{c}{{\bf ES-OV$_{\alpha}$ metric}} \\ \hline
Tuning parameters    & Percentage of           & Tributaries     & Sensitivities        & Specificities        \\
                     & correct classification  &                 &                      &                      \\  \hline          
$\alpha=0.5$ \& k=2  & $92.78\%$ ($3.25\%$)    & Anoia           & $95.77\%$($4.92\%$)  & $98.60\%$($2.14\%$)  \\
                     &                         & Cardener        & $85.25\%$($10.65\%$) & $97.06\%$($2.32\%$)  \\
                     &                         & Upper Llobregat & $93.93\%$($6.07\%$)  & $97.58\%$($2.34\%$)  \\
                     &                         & Lower Llobregat & $94.00\%$($6.48\%$)  & $97.24\%$($2.37\%$)  \\  \hline 
$\alpha=1$ \& k=3    & $89.88\%$ ($3.96\%$)    & Anoia           & $93.57\%$($5.92\%$)  & $97.17\%$($2.83\%$)  \\
                     &                         & Cardener        & $82.10\%$($12.82\%$) & $96.13\%$($2.83\%$)  \\
                     &                         & Upper Llobregat & $91.50\%$($7.42\%$)  & $96.42\%$($2.84\%$)  \\
                     &                         & Lower Llobregat & $89.88\%$($8.39\%$)  & $96.85\%$($2.63\%$)  \\  \hline \hline                    
\multicolumn{5}{c}{{\bf TC$_{\alpha}$ metric}} \\ \hline 
Tuning parameters    & Percentage of           & Tributaries     & Sensitivities        & Specificities        \\  
                     & correct classification  &                 &                      &                      \\  \hline            
$\alpha=0.35$ \& k=2 & $93.77\%$ ($3.13\%$)    & Anoia           & $96.73\%$($4.58\%$)  & $98.60\%$($1.83\%$)  \\
                     &                         & Cardener        & $87.80\%$($10.28\%$) & $97.66\%$($2.24\%$)  \\
                     &                         & Upper Llobregat & $94.18\%$($5.86\%$)  & $97.85\%$($2.30\%$)  \\
                     &                         & Lower Llobregat & $94.58\%$($6.18\%$)  & $97.65\%$($2.24\%$)  \\  \hline 
$\alpha=1$ \& k=2    & $86.55\%$ ($4.71\%$)    & Anoia           & $90.03\%$($7.41\%$)  & $95.99\%$($3.52\%$)  \\
                     &                         & Cardener        & $79.70\%$($13.45\%$) & $96.56\%$($2.66\%$)  \\
                     &                         & Upper Llobregat & $85.54\%$($8.84\%$)  & $95.95\%$($3.12\%$)  \\
                     &                         & Lower Llobregat & $89.08\%$($9.47\%$)  & $93.58\%$($3.69\%$)  \\  \hline \hline
\multicolumn{5}{c}{{\bf Aitchisonian metric}} \\ \hline
                     & Percentage of           & Tributaries     & Sensitivities        & Specificities        \\ 
                     & correct classification  &                 &                      &                      \\  \hline    
                     & $85.46\%$ ($5.07\%$)    & Anoia           & $87.40\%$($8.63\%$)  & $96.25\%$($2.94\%$)  \\
                     &                         & Cardener        & $77.65\%$($12.68\%$) & $95.91\%$($2.86\%$)  \\
                     &                         & Upper Llobregat & $89.89\%$($7.69\%$)  & $93.95\%$($3.75\%$)  \\
                     &                         & Lower Llobregat & $84.38\%$($9.88\%$)  & $94.49\%$($3.72\%$)  \\  \hline \hline              
\end{tabular}       
\caption{Classification results for the hydrochemical data. The number inside the parentheses indicates the standard error of the percentages.}
\label{tab1}
\end{center}
\end{small} 
\end{table}

\begin{figure}[!ht]
\centering
\begin{tabular}{cc}
\includegraphics[scale=0.37,trim=0 20 0 20]{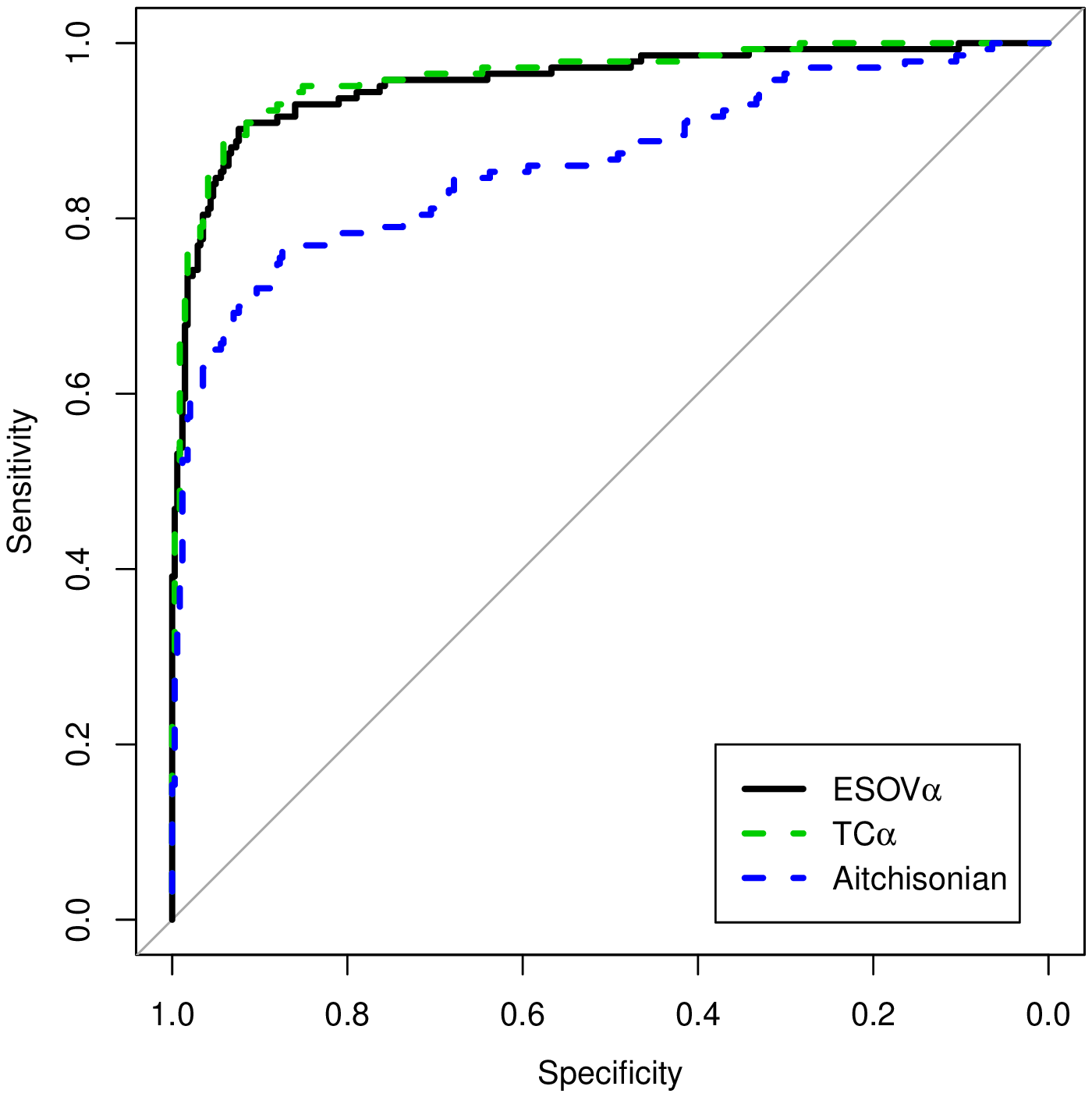} &
\includegraphics[scale=0.37,trim=0 20 0 20]{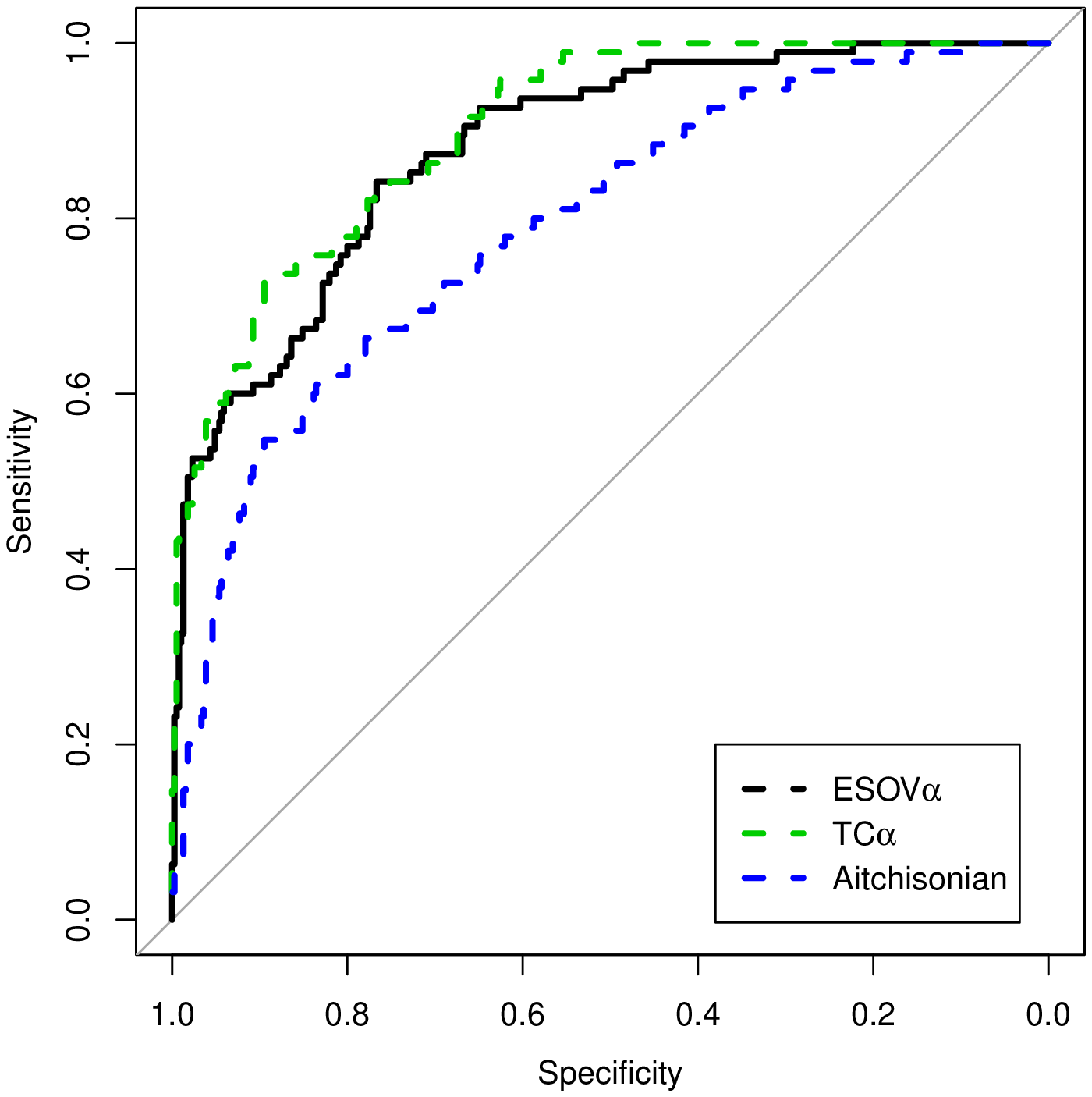} \\
\footnotesize{(a)}   &  \footnotesize{(b)}  \\
\includegraphics[scale=0.37,trim=0 20 0 20]{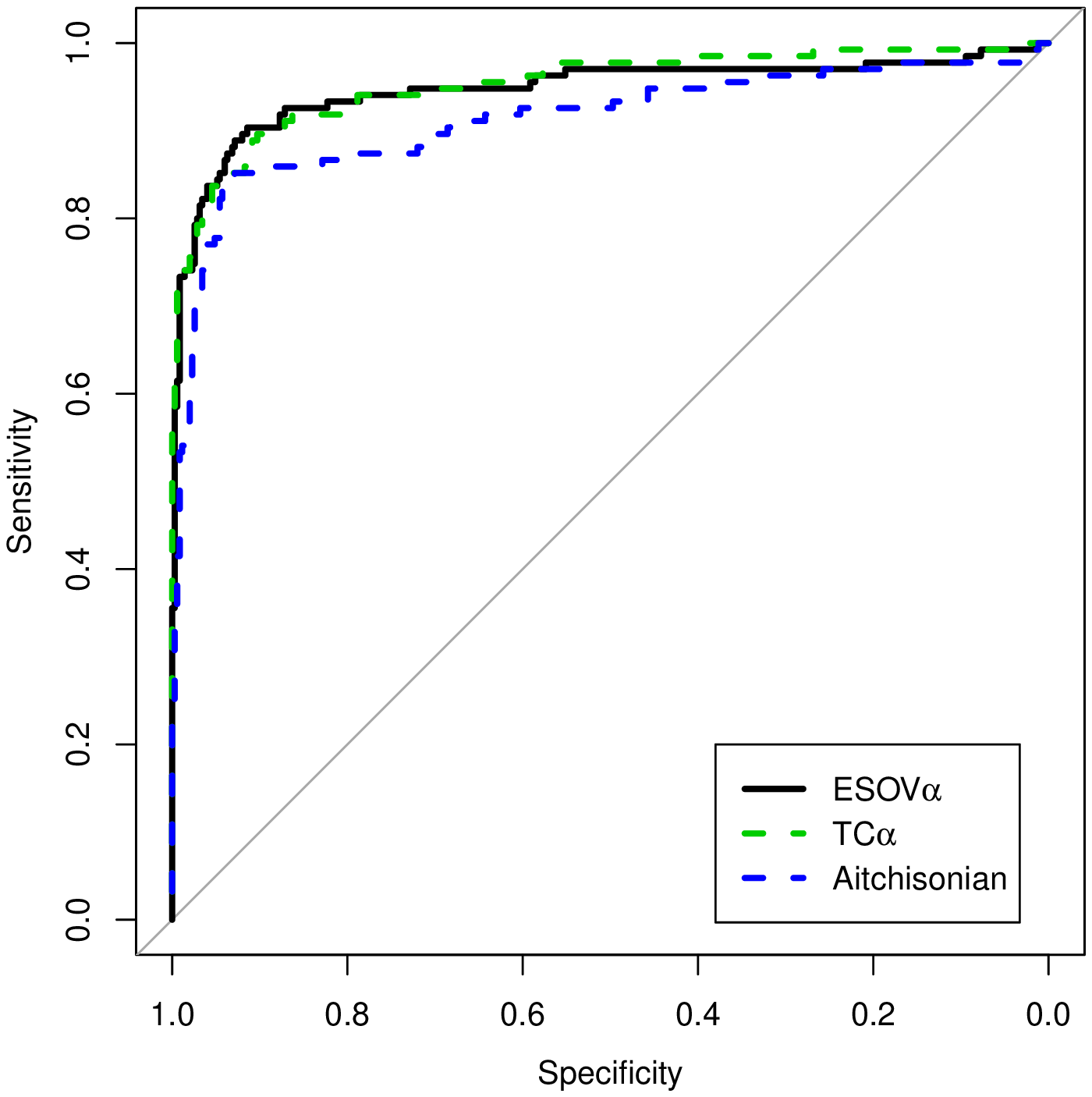} & 
\includegraphics[scale=0.37,trim=0 20 0 20]{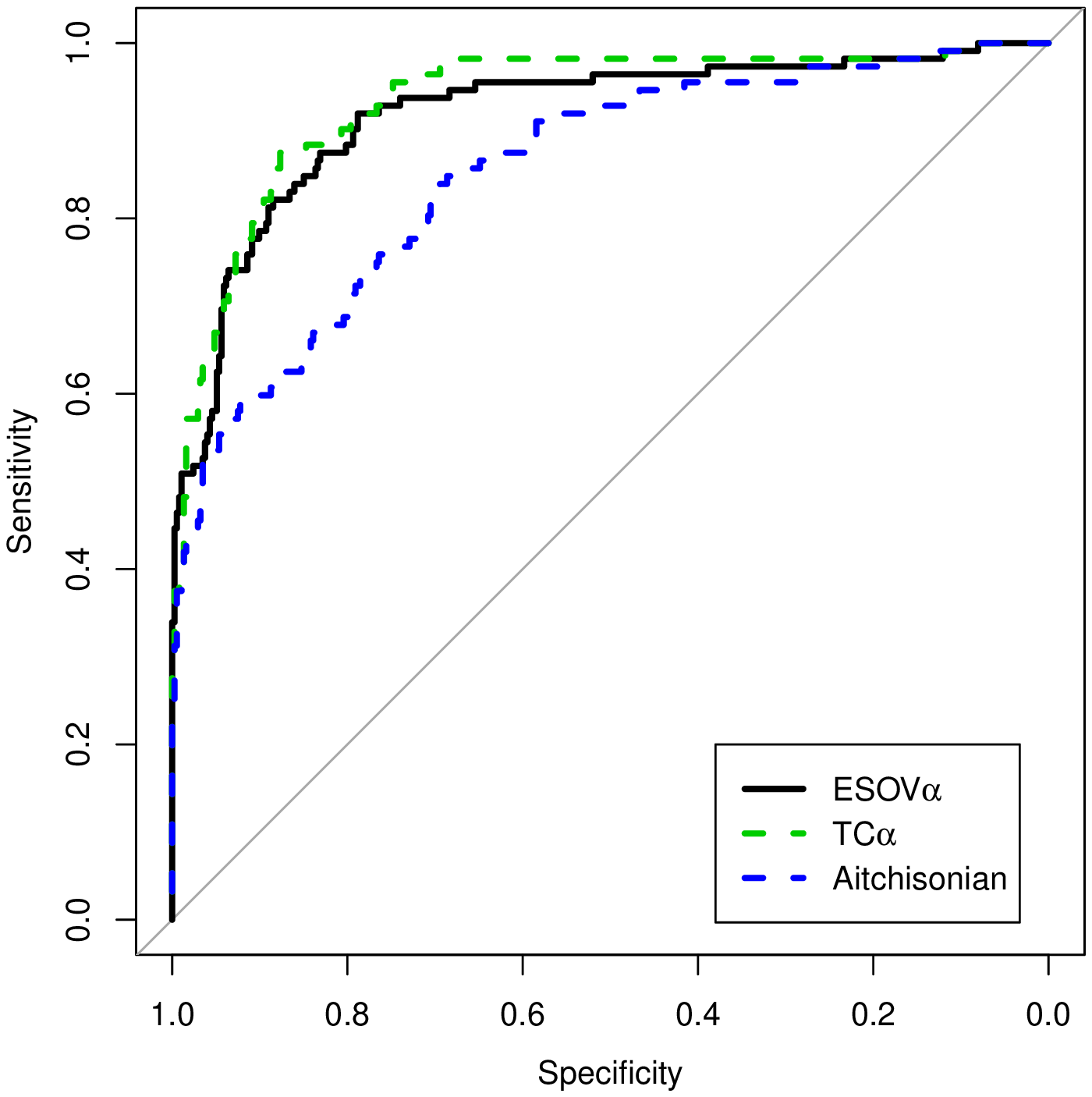} \\
\footnotesize{(c)}   &  \footnotesize{(d)}  \\
\end{tabular}
\caption{ROC curves for all tributaries using the three metrics For ES-OV$_{\alpha}$ we used $\alpha=0.35$ and $k=2$ and for TC$_{\alpha}$ we used $\alpha=0.5$ and $k=2$. For Each plot corresponds to one of the four tributaries (a) Anoia, (b) Cardener, (c) Upper Llobregat and (d) Upper Llobregat.}
\label{roc1}
\end{figure}

In addition we calculated the ROC curves for each of the three metrics. In order to do this we performed a $1$-fold cross validation. That is, we removed an observation and then using the parameters $\alpha$ and $k$ which are given in Table \ref{tab1} (since they produced the best results) we classified it. This procedure was repeated for all observations. Thus, we ended up with the predicted membership values for all observations based on the 3 metrics. This allowed us to draw the ROC curves for each tributary when all 3 metrics were used. The results are presented in Figure \ref{roc1}.

We can see that for all tributaries the ROC curves of the ES-OV$_{\alpha}$ metric (\ref{ESOVa}) and the TC$_{\alpha}$ metric (\ref{taxicaba}) are similar, whereas the ROC curve of the Aitchisonian metric (\ref{dist}) is always the lowest. 

\subsubsection*{\textit{Example 2. Forensic glass data with zero values}} 
In the second example we will use the forensic glass dataset which has $214$ observations from $6$ different categories of glass with $8$ chemical elements, in percentage form. The categories which occur are containers ($13$ observations), vehicle headlamps ($29$ observations), tableware ($9$ observations), vehicle window glass ($17$ observations), window float glass ($70$ observations) and window non-float glass ($76$ observations). This dataset contains a large number of zeros as well, thus excluding LRA from being applied here. The data are available from the \href{http://archive.ics.uci.edu/ml/datasets/Glass+Identification}{UC Irvine Machine Learning Repository}.  

\begin{figure}[!ht]
\centering
\begin{tabular}{cc}
\includegraphics[scale=0.45,trim=0 20 0 20]{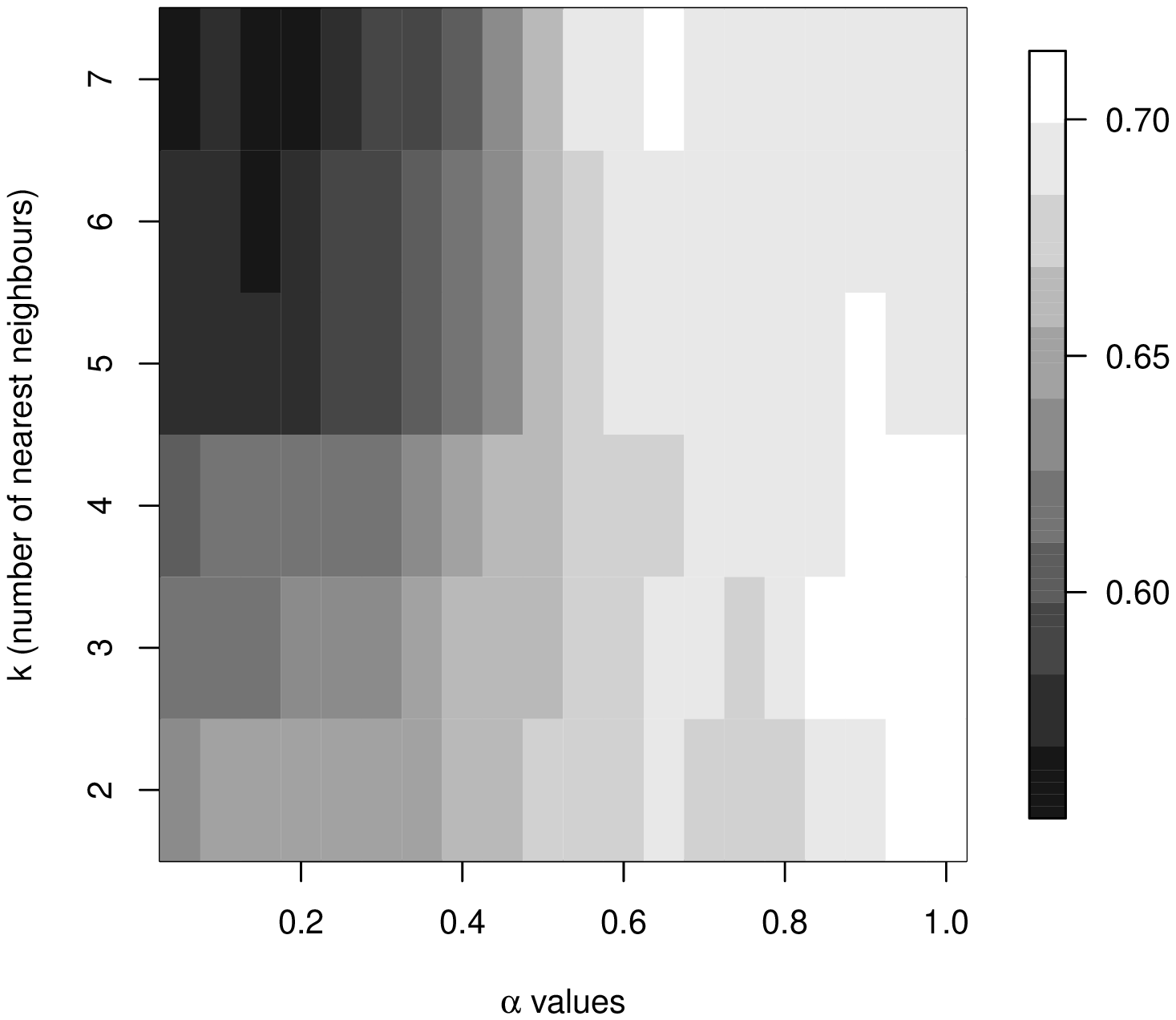} &
\includegraphics[scale=0.45,trim=0 20 0 20]{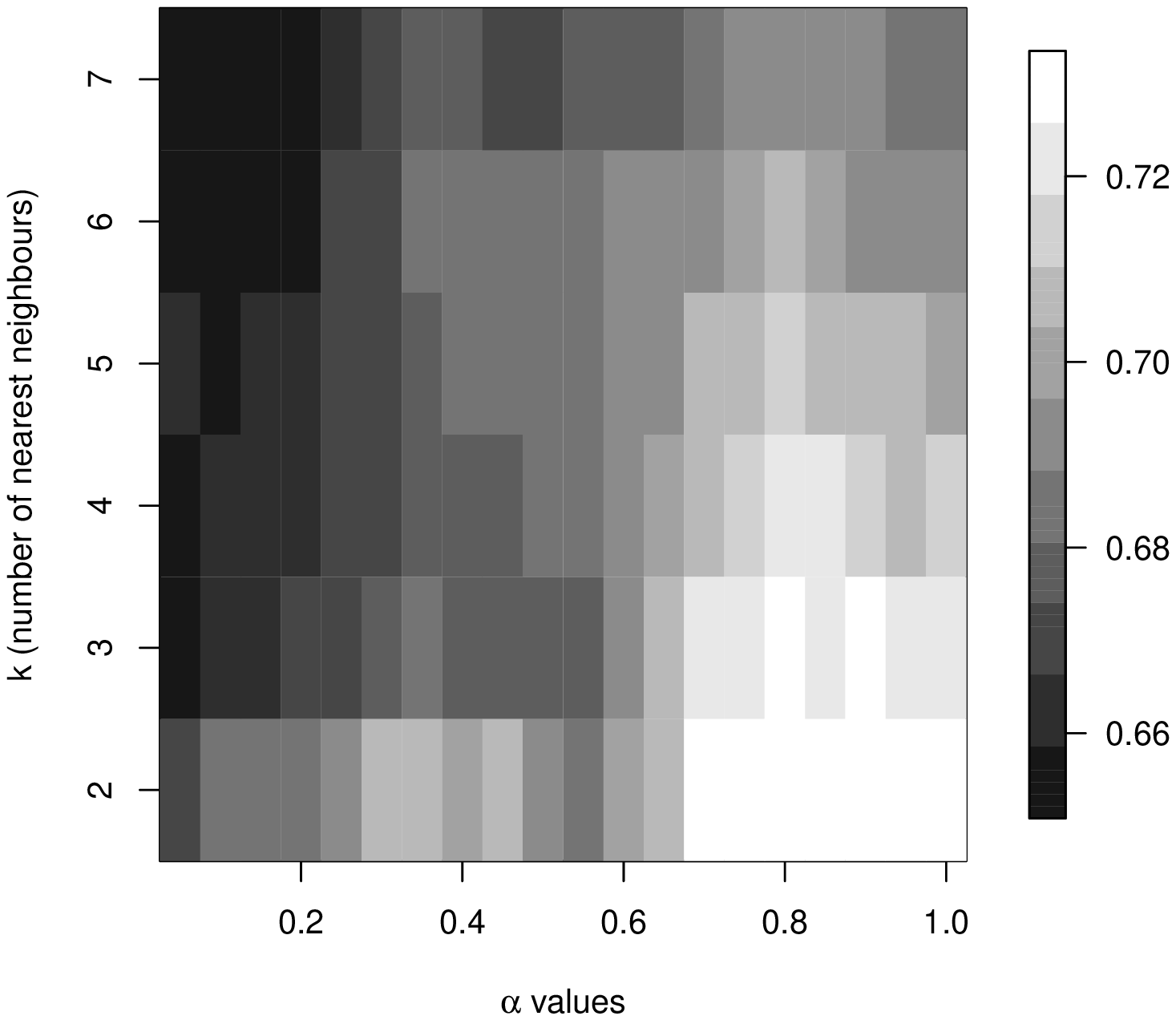} \\
\footnotesize{(a)}   &  \footnotesize{(b)}  
\end{tabular}
\caption{The estimated percentage of correct classification for the forensic glass data as a function of $k$, the nearest neighbours and of $\alpha$ using the (a) ES-OV$_{\alpha}$ metric (\ref{ESOVa}) and (b) TC$_{\alpha}$ (\ref{taxicaba}).} 
\label{fig7}
\end{figure}

An interesting feature of this dataset is that it contains many zero values. This means that the Aitchisonian metric (\ref{dist}) is not to be used. The ES-OV$_{\alpha}$ and the TC$_{\alpha}$ metrics on the other hand are not affected by the presence of zeros, since $0\log0=0$. In this example the sample size of the test data was equal to $30$, hence we used $184$ compositional vectors to train the $k$-NN algorithm. Again, the test data were chosen via stratified random sampling to avoid having categories not been selected in the test sample. Figure \ref{fig7} shows the estimated percentage as a function of $k$ and $\alpha$ using both metrics.   

This is a simpler case to draw conclusions, since the best results are obtained when $\alpha=1$ and $k=2$ for both metrics, thus the ES-OV (\ref{ESOV}) and the TC (\ref{taxicab}) metrics should be used, with the estimated percentage of correct classification being $71.45\%$ and $73.35\%$ respectively. Table \ref{tab2} presents analytical information of the classification results. Estimates of the sensitivities and of the specificities for each category of glass are also given. 

The mean sensitivities of ES-OV$_{\alpha}$ metric (\ref{ESOVa}) for Tableware and Vehicle window are low and the same is true for the Vehicle window when TC$_{\alpha}$ (\ref{taxicaba}) is used. We observed that many times, Tableware and Vehicle window were being wrongly classified as Vehicle float. A possible reason for this could be the small sample size of Tableware (this type of glass had the minimum number of observations). A chemist or a forensic scientist could perhaps give a possible answer to this (if that is the case of these types of glass being of similar structure).   

\begin{table}[h]
\begin{small}
\begin{center}
\begin{tabular}{ccccc} \hline
\multicolumn{5}{c}{{\bf ES-OV$_{\alpha}$}} \\ \hline
Tuning parameters    & Percentage of           & Glass             & Sensitivities         & Specificities        \\
                     & correct classification  & categories        &                       &                      \\  \hline          
$\alpha=1$ \& k=3    & $71.45\%$ ($7.76\%$)    & Containers        & $77.25\%$($29.57\%$)  & $97.46\%$($2.76\%$)  \\
                     &                         & Vehicle headlamps & $80.88\%$($16.99\%$)  & $96.44\%$($3.60\%$)  \\
                     &                         & Tableware         & $36.50\%$($48.26\%$)  & $96.95\%$($3.00\%$)  \\
                     &                         & Vehicle window    & $29.25\%$($31.81\%$)  & $97.50\%$($2.97\%$)  \\  
                     &                         & Vehicle float     & $81.65\%$($11.60\%$)  & $82.30\%$($7.28\%$)  \\
                     &                         & Non-window float  & $68.55\%$($13.39\%$)  & $90.50\%$($6.30\%$)  \\  \hline \hline                    
\multicolumn{5}{c}{{\bf TC$_{\alpha}$}} \\ \hline 
Tuning parameters    & Percentage of           & Glass             & Sensitivities         & Specificities        \\  
                     & correct classification  & categories        &                                              \\  \hline            
$\alpha=1$ \& k=3    & $73.35\%$ ($8.00\%$)    & Containers        & $77.75\%$($30.37\%$)  & $98.18\%$($2.43\%$)  \\
                     &                         & Vehicle headlamps & $82.62\%$($16.66\%$)  & $99.15\%$($1.77\%$)  \\
                     &                         & Tableware         & $74.50\%$($43.70\%$)  & $98.14\%$($2.70\%$)  \\
                     &                         & Vehicle window    & $29.75\%$($31.74\%$)  & $95.48\%$($3.71\%$)  \\  
                     &                         & Vehicle float     & $77.90\%$($12.58\%$)  & $82.10\%$($7.52\%$)  \\
                     &                         & Non-window float  & $72.86\%$($14.45\%$)  & $90.11\%$($6.99\%$)  \\  \hline \hline             
\end{tabular}       
\caption{Classification results for the forensic glass data. The number inside the parentheses indicates the standard error of the percentages.}
\label{tab2}
\end{center}
\end{small} 
\end{table}

\begin{figure}[!ht]
\centering
\begin{tabular}{ccc}
\includegraphics[scale=0.33,trim=0 20 0 20]{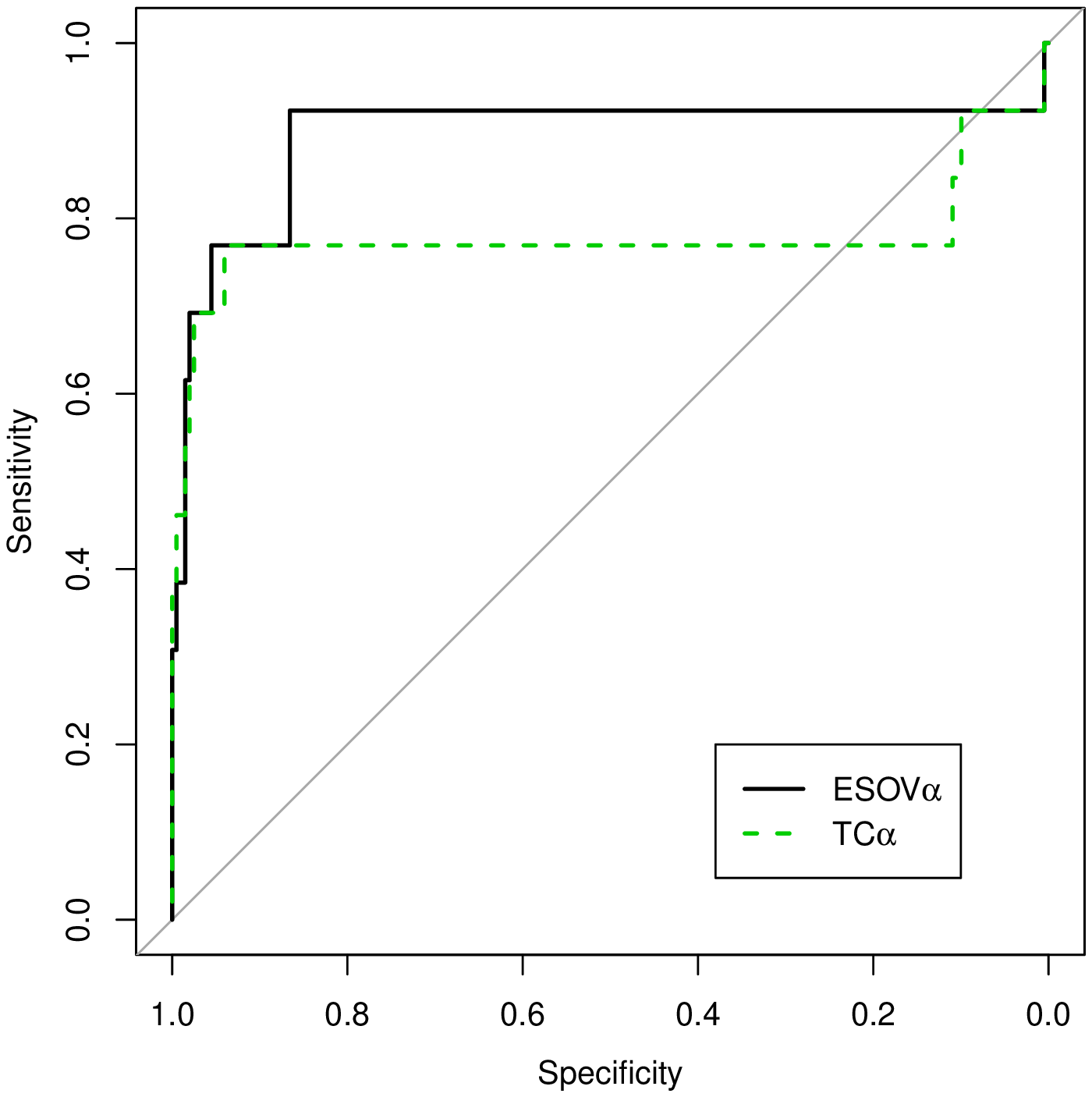} &
\includegraphics[scale=0.33,trim=0 20 0 20]{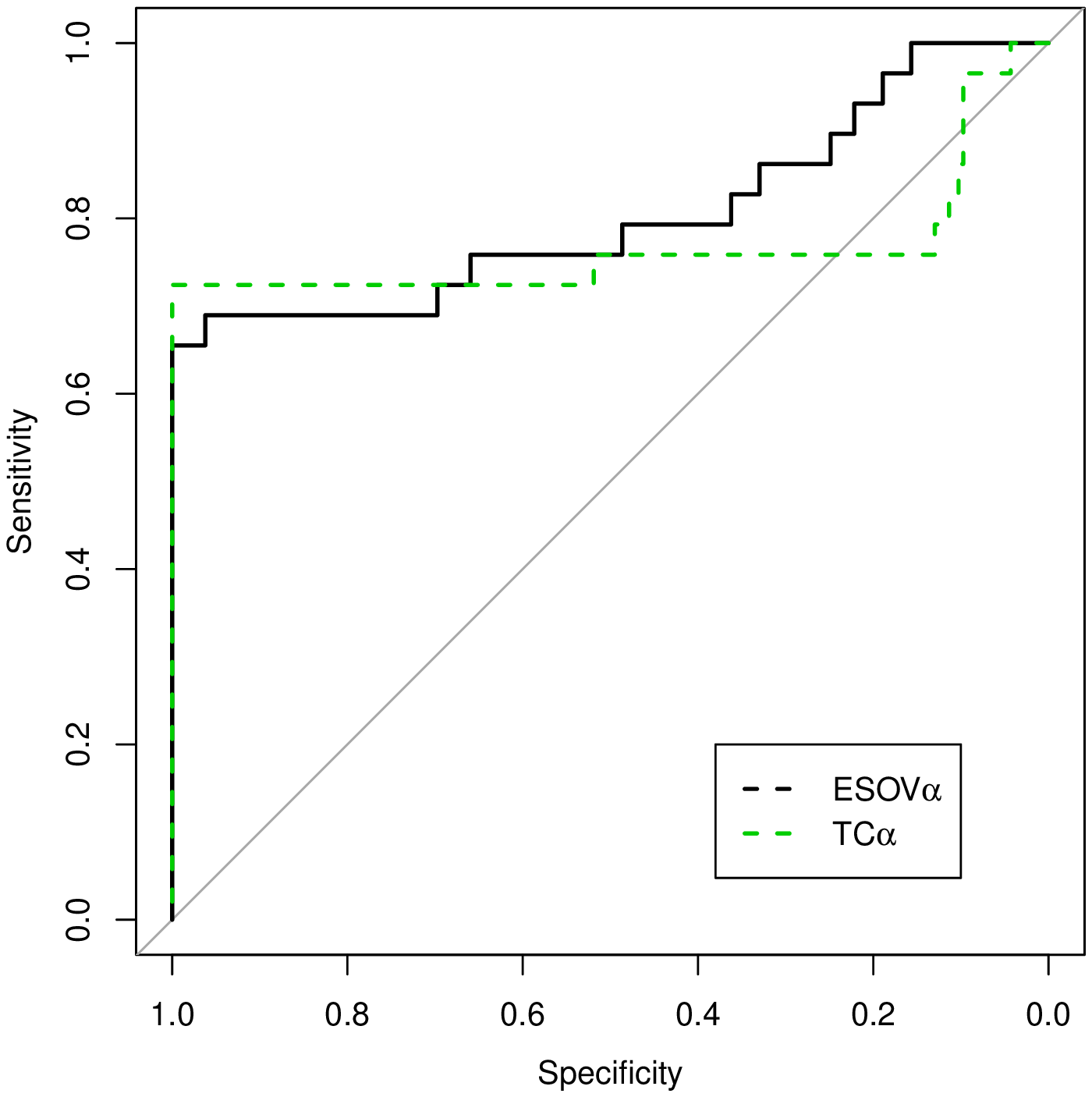} &
\includegraphics[scale=0.33,trim=0 20 0 20]{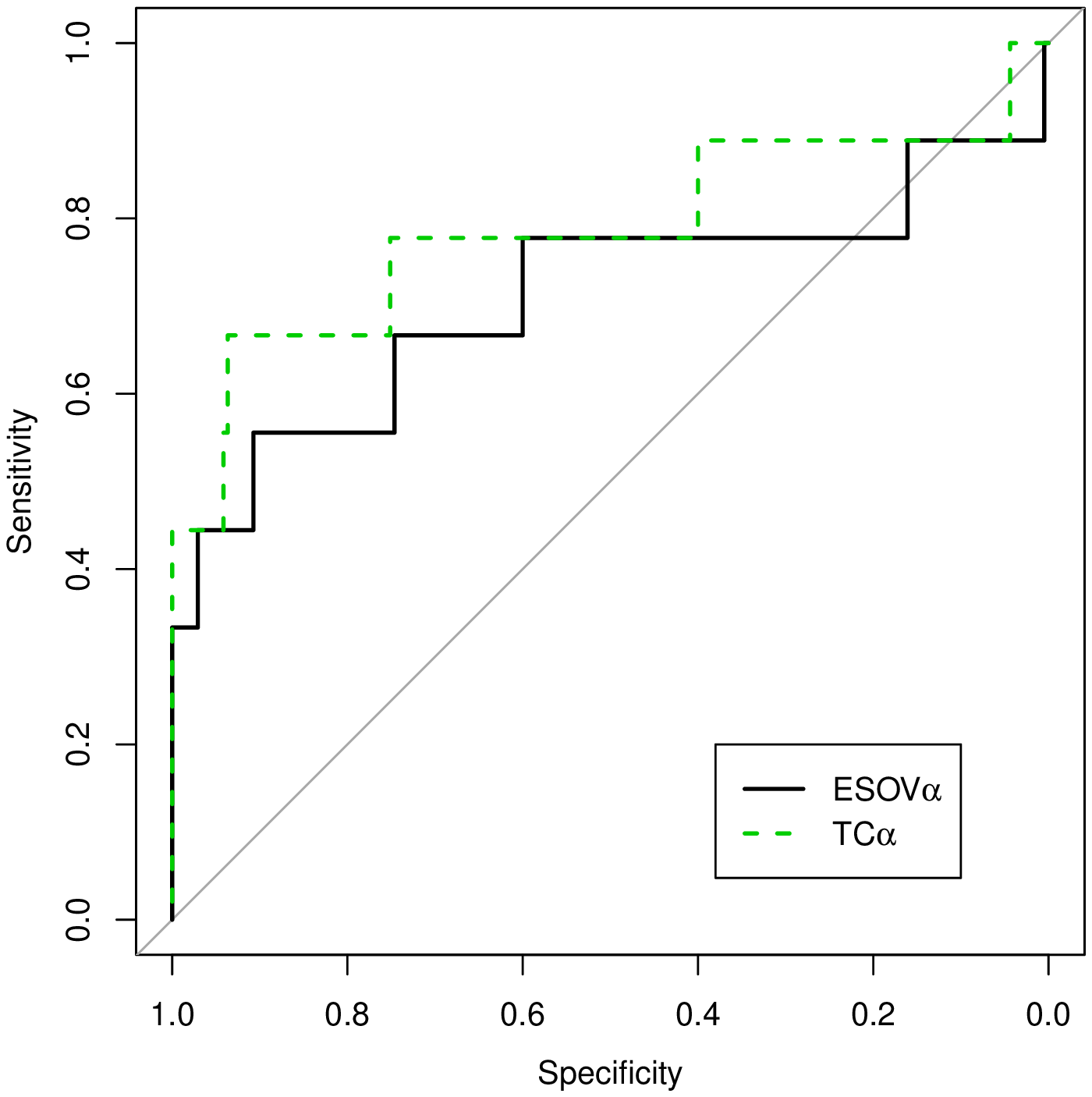}  \\
\footnotesize{(a)}   &  \footnotesize{(b)}   &  \footnotesize{(c)} \\
\includegraphics[scale=0.33,trim=0 20 0 20]{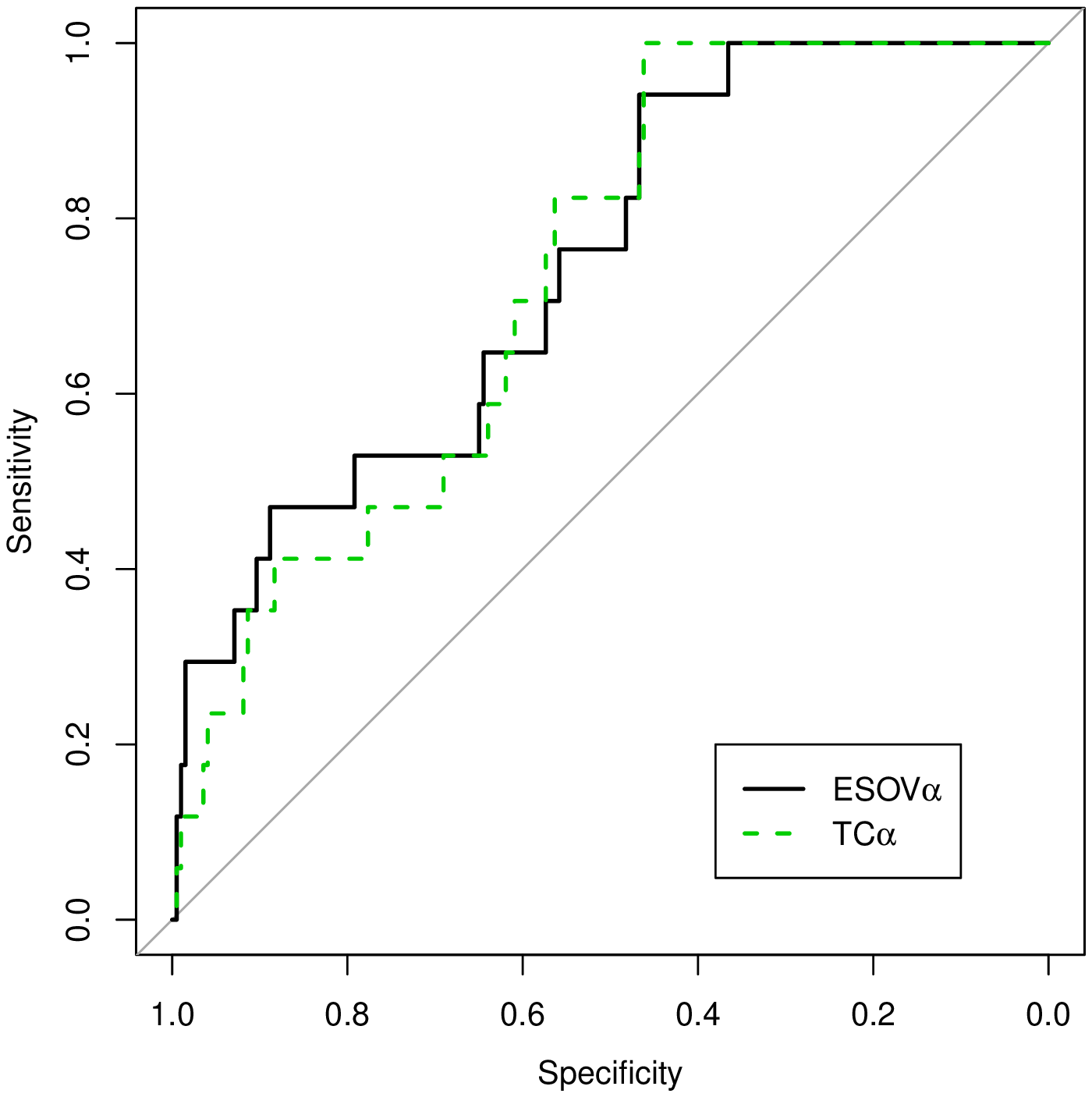} &
\includegraphics[scale=0.33,trim=0 20 0 20]{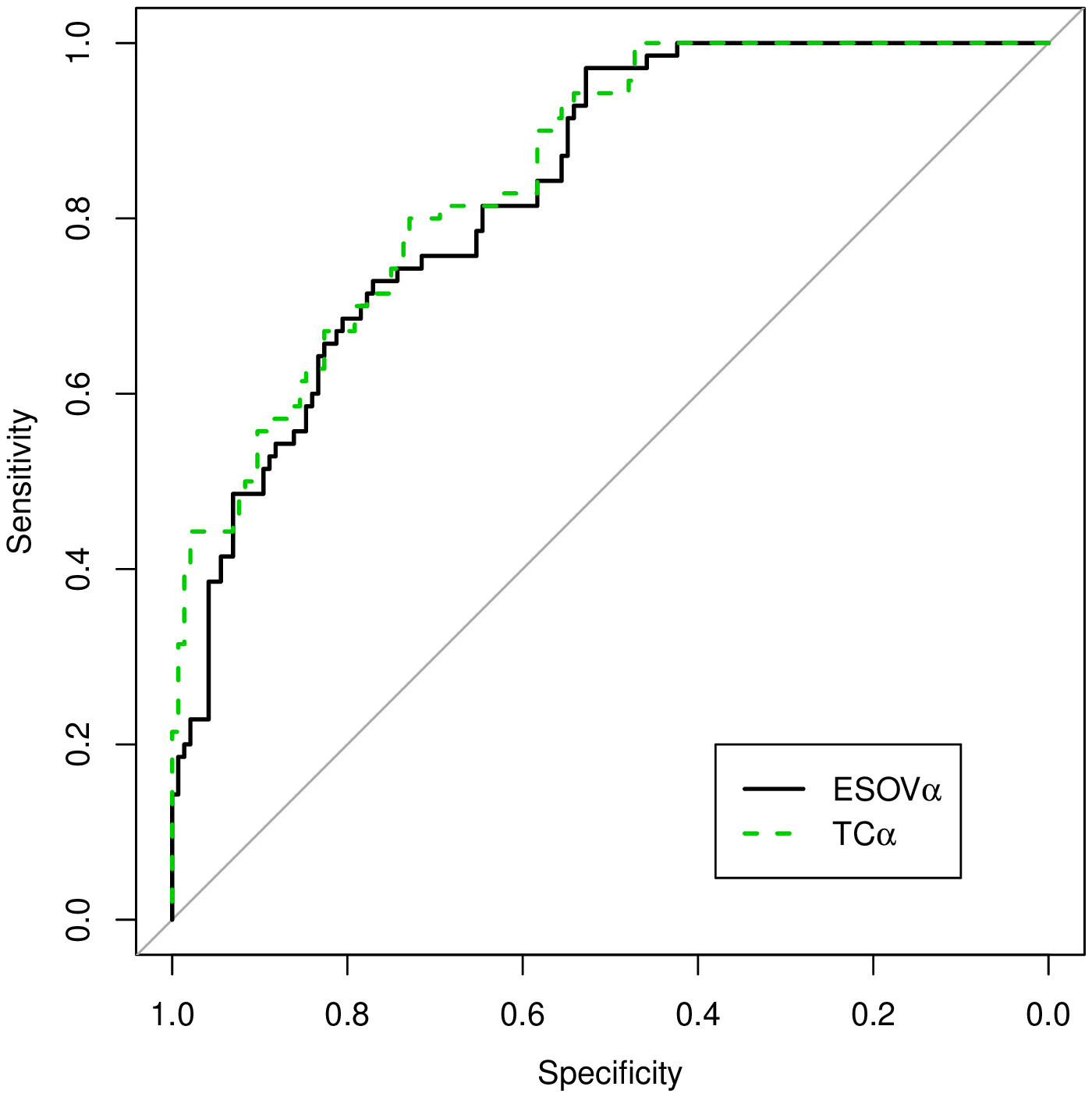} &
\includegraphics[scale=0.33,trim=0 20 0 20]{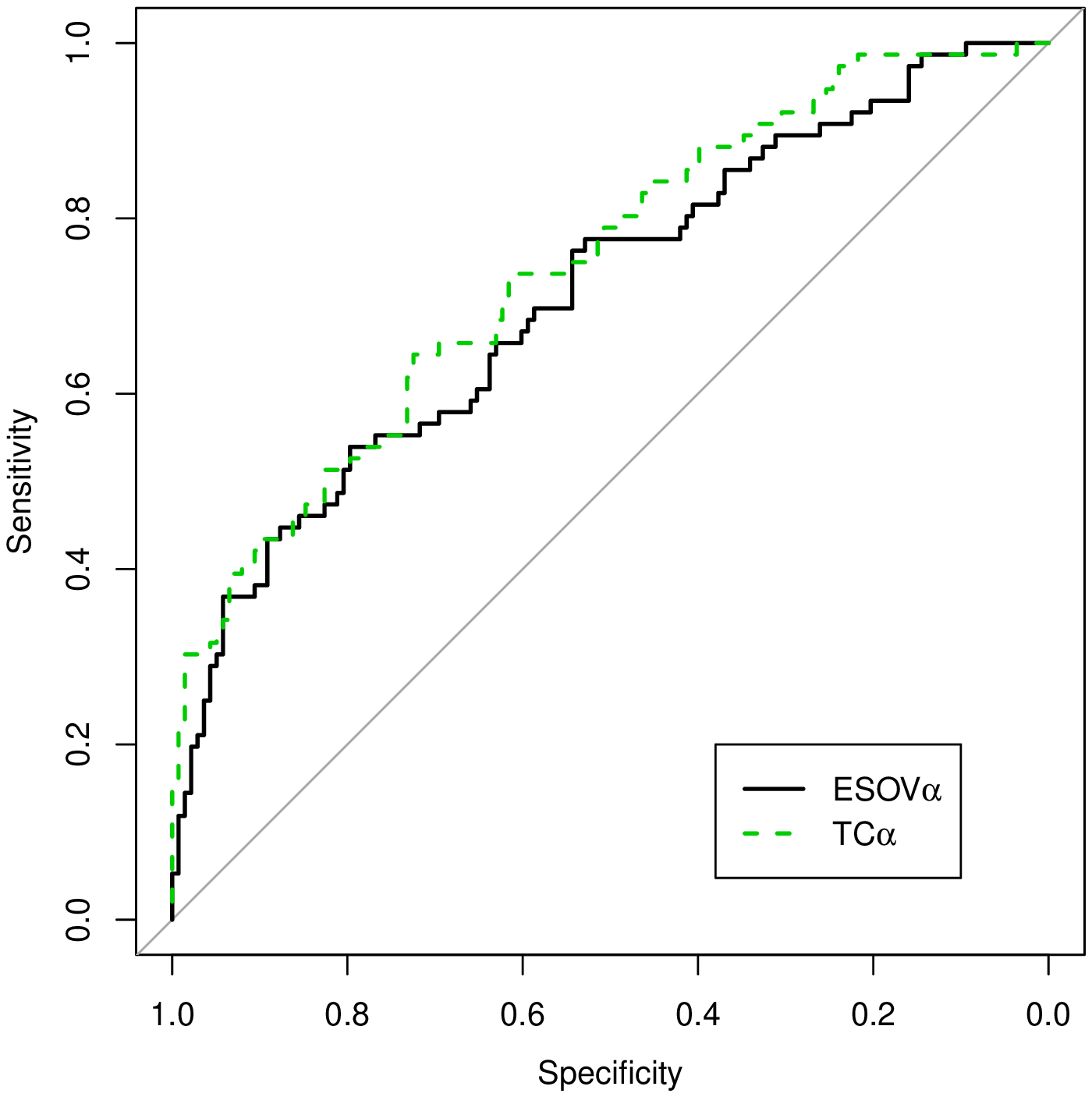}  \\
\footnotesize{(d)}   &  \footnotesize{(e)}   &  \footnotesize{(f)} 
\end{tabular}
\caption{ROC curves for all tributaries using the three metrics In all cases $\alpha=1$ and $k=3$ were used in both metrics. Each plot corresponds to one of the six glass categories  (a) containers, (b) vehicle headlamps, (c) tableware, (d) vehicle window glass, (e) window float glass and (f) window non-float glass.}
\label{roc2}
\end{figure}

The ROC curves for each glass category (based on $1$-fold cross validation) using both metrics are presented in Figure \ref{roc2}. We cannot say that one metric does better than the other always. For some glass categories, the two ROC curves are similar and for some others one seems a bit better than the other. 

\section{Conclusions}
We suggested the use of a recently developed metric (\ref{ESOV}), for supervised classification when the $k$-NN algorithm is implemented. We also added a free parameter to the metric with the intention of improving the classification results. This free parameter was used to generalize the taxicab metric as well. The examples showed that both the ES-OV$_{\alpha}$ (\ref{ESOVa}) and the taxicab$_{\alpha}$ (\ref{taxicaba}) metric can be used for supervised clustering of compositional data, but can also be used in other scenarios as well.  

An advantage of both metrics over the Aitchisonian metric (\ref{dist}) is that they handle zeros naturally. This implies that no zero value replacement is necessary either parametrically \citep{martin2012} or non parametrically \citep{ait2003}. In order to appreciate the importance of this advantage one can think of large datasets with many zeros. 

The two metrics outbalanced the Aitchisonian metric (\ref{dist}) in the examples presented in this manuscript. When it comes  to comparing the the ES-OV$_{\alpha}$ (\ref{ESOVa}) and the taxicab$_{\alpha}$ (\ref{taxicaba}) metric between them we cannot say one is better than the other. 

A closer examination of the ROC curves revealed valuable information, especially for the FGL data example (where zeros are present) regarding the classification abilities of the ES-OV$_{\alpha}$ (\ref{ESOVa}) and the taxicab$_{\alpha}$ (\ref{taxicaba}) metric. The sensitivities and specificities revealed interesting patterns of the misclassification rates not captured by the percentage of correct classification. In addition, the ROC curves provided graphical evidence as for the ability of each metric to classify the observations. 

\section*{Acknowledgements}
The author would like to acknowledge the anonymous referee for pointing out very interesting details.

\end{document}